\documentclass[twocolumn]{aastex631}
\usepackage{amsmath}
\usepackage{graphicx}
\usepackage[caption=false]{subfig}
\usepackage{float}
\usepackage{multirow,array,booktabs}
\usepackage{bm}

\newcommand{\Msol}{$M_\odot$}
\DeclareUnicodeCharacter{2212}{-}

\newcommand{\ucsdp}{Department of Physics, University of California, San Diego, La Jolla, CA 92093}
\newcommand{\ucsda}{Department of Astronomy and Astrophysics, University of California, San Diego, La Jolla, CA 92093}
\newcommand{\ufl}{Department of Astronomy, University of Florida, Gainesville, FL 32611
}
\newcommand{\ncu}{Centre for Astronomy, Faculty of Physics, Astronomy and Informatics, Nicolaus Copernicus University, Grudziadzka 5, PL-87-100 Torun, Poland}
\begin{document}
\title{On The Orbital Evolution of Multiple Wide Super-Jupiters: How Disk Migration and Dispersal Shape the Stability of The PDS 70 System}

\author[0000-0001-5173-2947]{Clarissa R. Do Ó}
\affiliation{\ucsdp}

\author{Jaehan Bae}
\affiliation{\ufl}

\author[0000-0002-9936-6285]{Quinn M. Konopacky}
\affiliation{\ucsda}

\author[0000-0002-9242-9052]{Jayke S. Nguyen}
\affiliation{\ucsda}

\author{Patrick Diamond}
\affiliation{\ucsdp}
\affiliation{\ucsda}

\author{Krzysztof Goździewski}
\affiliation{\ncu}

\author{Dawid Jankowski}
\affiliation{\ncu}
\begin{abstract}
Direct imaging has revealed exoplanet systems hosting multiple wide-orbit Super-Jupiters, where planet-planet interactions can shape their long-term dynamical evolution. These strong perturbations may lead to orbital instability, raising questions about the long-term survival of such systems. Shortly after formation, planet-disk interactions can shepherd planets into mean-motion resonances, which may promote long-term stability as seen in HR 8799. However, early-stage processes such as disk photoevaporation and viscosity can influence these outcomes. The $\sim$5 Myr-old PDS 70 system offers a unique laboratory to investigate these processes: its two massive ($>$4 $M_{Jup}$), wide-orbit ($>$20 AU) giants are still embedded in their natal disk. We perform 2D hydrodynamic simulations of the system, allowing the disk to disperse via photoevaporation. Once the disk dissipates, we continue to track the planets' orbital evolution over Gyr timescales using N-body simulations. We find that the system is likely to remain stable for $>$ 1 Gyr. To assess the importance of disk-driven evolution, we compare these results with disk-free N-body simulations using orbital parameters constrained by orbit fits that include recent relative astrometry and radial velocities from the literature. In this case, we find that only $\lesssim 4\%$ of posterior is stable for 100 Myr, highlighting the importance of considering disk-driven evolution for long-term dynamics stability of exoplanetary systems. We also simulate two three-planet configurations including the proposed inner candidate “PDS 70 d”, finding that a higher photoevaporation leads the system to become unstable in $<$ 10 Myr.

\end{abstract}

\section{Introduction} \label{intro}

Direct imaging with large ground-based telescopes has enabled the detection of a population of young ($<$ 1 Gyr), widely separated ($>$ 5 AU) gas giant exoplanets ($>$ 2 $M_{Jup}$). Most of these systems are within their first tens of Myr of age. Because they have recently formed, they are ideal for comparing planet formation models with observations, as their formation pathways can be traced by observables such as luminosity, separation, and orbital architecture. After formation in a protoplanetary disk, planets interact gravitationally with the disk, which can affect their orbital evolution via inward/outward migration \citep{Kley_Nelson_2012, Bitsch_Kley_2010} and resonance locking in multi-planet systems \citep{Masset2001, Snellgrove2001, Bae2019}. These interactions also lead the disk to form structured gaps, rings and spirals \citep{Isella2016, Andrews2016, Bae2016, Bae2017}.
\par
The Atacama Large Millimeter/submillimeter Array (ALMA) has recently imaged systems with protoplanetary disks by detecting the emission from cold dust in the disk at sub-mm/mm wavelengths (e.g. \citealt{ALMA2014, Andrews2016, Andrews2018}). These images allow for the detailed study of planet-disk interactions \citep{Bae2023}. ALMA has successfully imaged dozens of these disks, with only one system having confirmed the detection of the protoplanets themselves (PDS 70; \citealt{Keppler2018, Haffert2019}).
The sample size is limited by the short duration of the gas disk, which typically lasts a few Myr \citep{Ribas2015}, and by instrument detection limits \citep{Benisty2023}. 
\par
PDS 70's inner planet, PDS 70 b, was initially detected via multi-band thermal emission \citep{Keppler2018, Muller2018} and H-$\alpha$ emission, with the outer planet, PDS 70 c, subsequently detected \citep{Wagner2018, Haffert2019} near the inner edge of the outer disk. Several works have aimed to characterize the system, including detections of circumplanetary disks, accretion signatures \citep{Haffert2019, Close2025}, and dust structures \citep{Bae2019}. The separation of the planets suggest they are near a 2:1 mean-motion resonance, which is believed to be a consequence of planet-planet and planet-disk interactions shortly after formation \citep{Masset2001, Snellgrove2001, Batygin2015}. For this reason, PDS 70 has become an important testbed not only for understanding the processes of planet formation, but also for observing the early stages of planetary system evolution in real time.
\par
Questions about orbital evolution that can be probed using PDS 70 include: How does the resonance locking occur? Is the orbital resonance necessary for the long-term survival of multiple massive gas giants? Can we expect dynamical stability over Gyr timescales given the system’s youth? As new direct imaging instruments become more sensitive to older planets, constraining the orbital evolution of their youthful counterparts becomes increasingly important. In this work, we model the possible orbital outcomes of the PDS 70 protoplanets using two-dimensional hydrodynamic simulations coupled with N-body dynamics. \par
To explore the long-term orbital evolution of PDS 70’s protoplanets, we must account for important physical processes that can shape young planetary systems: photoevaporative disk dispersal, planet-disk and planet-planet interactions and disk viscosity. Below we describe these important processes that have been investigated in previous works. We motivate the evaluation of the PDS 70 system using all of these processes acting in concert to assess its long-term orbital evolution. We then aim to compare the results to the traditionally used orbit fit posteriors coupled with N-body without considering disk migration.
\par
\subsection{Disk Photoevaporation}
We consider the evolution of the system as the gas disk undergoes photoevaporation due to heating from the central star, a process that controls the duration of planet-disk interactions during the final stages of the disk’s lifetime \citep{Rab2016}. There are two reasons why we include photoevaporation. First, it is an important process that has not been widely explored in the context of multi-planet systems found with direct imaging (including this system; e.g., \citealt{Bae2019, Toci2020}), even though it plays a major role in driving disk dissipation \citep{Owen2012} and halting planet migration. Planet migration may explain the orbital configurations observed in older, gas disk-free systems such as HR 8799 \citep{Gozdziewski2014, Zurlo2022}. Second, in order to assess this system on longer timescales ($>$ 100 Myr), it is important to consider that the disk will not last for the entirety of the system's existence, but rather just a few Myr \citep{Ribas2015}. In order to make a true comparison to older systems ($>$ 10 Myr) found with direct imaging, where the gas disk is no longer present, it is important to invoke a process that will dissipate the disk throughout the system's lifetime. The internal photoevaporation of the gas disk is the dominant form of disk dispersal after most of the viscous accretion has occurred \citep{Champion2019}. Different approaches can model this process, which is caused by ionizing photons in the extreme ultra-violet (EUV) and X-ray. \par
\citealt{Owen2012} demonstrated that the X-ray component of the star's radiation dominates the mass loss by effectively heating the gas in the inner disk. This effectiveness can be attributed to the larger penetration depth of X-ray photons. The modeling of the photoevaporation process agrees with the observations of disk lifetimes; \citealt{Ribas2015} found that the majority of disks have a lifetime of up to $\sim$ 10 Myr. Recent hydrodynamic works have attempted to constrain how the X-ray radiation affects the density of protoplanetary disks. The general formulation involves calculating the temperature and ionization structure of a disk around a T Tauri star irradiated by X-ray photons (\citealt{Ercolano2008a}; \citealt{Ercolano2008b}; \citealt{Owen2010}; \citealt{Picogna2019}; \citealt{Sellek2022}). The X-ray flux is usually obtained from a synthetic spectrum which varies with X-ray luminosity. Then, the photons are allowed to be absorbed, re-emitted and scattered by the material in the disk. 
\par
Many of the currently directly imaged companions orbit young stars, which are known to be more active than main-sequence stars (e.g. \citealt{Feinstein2020}, \citealt{Johnstone2021}). For that reason, in addition to the inherent disk parameters, the photoevaporation due to the host star's radiation can halt planet migration and this can have important consequences on the system's evolution.
\subsection{Planet-Disk and Planet-Planet Interactions}
The presence of a planet generates a perturbation in the disk structure due to its gravitational potential, which leads to spiral density waves launched at Lindblad resonance locations \citep{GoldreichTremaine1980} and density asymmetries in the planet's co-rotation region \citep{Ward1992}. The planet interacts with the disk, leading to significant exchange in angular momentum between the two and a subsequent migration of the planet. Different configurations of the disk and planet can lead to inward or outward migration (e.g. \citealt{Papaloizou2000, Papaloizou2001,Bitsch2010,Kley2012}). Very massive planets ($>$ 1 $M_{Jup}$) can also carve gaps in their protoplanetary disks due to a torque imbalance between the Lindblad torques caused by the planet, and the viscous torques, which refill the material in the co-rotation region \citep{FungShiChiang2014}. This leads to gas being repelled from the local vicinity of the planet \citep{LinPapaloizou1993}. \par
If there are two massive planets in the disk, convergent migration can lead to resonance locking due to the creation of a common gap \citep{Snellgrove2001}. The dominance of these different processes can vary significantly depending on specific disk conditions and planet location/masses, making it difficult to construct a robust analytical model for a specific case. Regardless, the relative planet masses can dictate whether the final system will migrate outwards or inwards, as was stipulated for the formation of the RV-detected GJ 876 System \citep{Masset2001} and our Solar System in the Grand Tack Model \citep{Walsh2011}. 
\par
\subsection{The viscosity parameter $\alpha$}
Most protoplanetary disk evolution thus far has been modeled using the viscous accretion prescription \citep{ShakuraaSunyaev1973}. This formulation relies on a viscosity equation that requires a viscosity ``efficiency" parameter $\alpha$, which describes how efficiently angular momentum and mass are redistributed in the disk. The origin of this transport can have a variety of causes, such as vertical shear instability and non-ideal MHD effects (see \citealt{Lesur2023} and references therein). Despite being a useful and simplified parameter, $\alpha$ is difficult to constrain observationally. Constraining $\alpha$ observationally may be possible by reproducing the dust distribution in an observed disk \citep{Pinte2016} or measuring stellar accretion rates \citep{Rafikov2017}. Current values vary by 2 orders of magnitude in simulations ($10^{-4} - 10^{-2}$); \citealt{Bae2019, Toci2020, Thanathibodee2020, Joyce2023, Hartmann1998, Rafikov2017, Sellek2020}). Such a variation will cause different migration outcomes for the PDS 70 protoplanets, as the viscosity dictates if a gas giant protoplanet will carve a gap in its disk, and how wide and deep this gap is. Since the formation of the gap affects the planet migration direction \citep{Crida2006, Afkanpour2024}, and disk viscosity values dictate this gap shape and width \citep{Kanagawa2015}, it is important to consider a range of $\alpha$ values for the PDS 70 disk. 
\par

\subsection{The PDS 70 System: General Properties}
The PDS 70 system is a young (5.4 Myr), directly imaged planetary system located about 110 pc from the Earth. It hosts a highly structured transitional disk and two confirmed protoplanets, named PDS 70 b and PDS 70 c, which are at a de-projected distance of $\sim$ 20 and 35 AU from the K7-type host star \citep{Keppler2018}. These distances are near a 2:1 resonance location (e.g. \citealt{Bae2019}; \citealt{Wang2021}). Table \ref{tbl:pds70properties} presents the PDS 70 system's general properties. \par
In this work, we simulate the PDS 70 system using hydrodynamics coupled with N-body simulations. We implement a photoevaporation prescription such that the gas disk dissipates throughout the hydrodynamic integration period. Once the gas disk is fully dissipated, we employ N-body simulations to analyze the system's long-term orbital evolution.
\begin{deluxetable}{ccc} [ht!]
\tablecaption{PDS 70 System Properties} \label{tbl:pds70properties}
\tablewidth{20pt}
\tablecolumns{3}
\tabletypesize{\scriptsize}
\tablehead{\colhead{Property} & \colhead{Value} & \colhead{Reference}}

\startdata
Age & 5.4 $\pm$ 1.0 Myr & 1\\ 
Distance & 113.43 ± 0.52 pc & 1 \\
$M_{*}$ & 0.85 \Msol & 2 \\
Spectral Type & K7 & 1\\
$\overline{L_x}$ & $1.37 $x$10^{30}$ $ergs~s^{-1}$  & 3  \\
$\dot{M_{*}}$ & $10^{-10}$\Msol $~yr^{-1}$ & 3 \\
$M_b$ & 4 -- 17 $M_{Jup}$ & 4, 5, 6\\
$M_c$ & 4 -- 12 $M_{Jup}$ & 5, 6\\
$\dot{M_{b}}$ and $\dot{M_{c}}$ & $10^{-8} -10^{-7} M_{Jup}$ $yr^{-1}$ & 3, 7 \\
\enddata
\tablecomments{References: (1) \citealt{Keppler2018}; (2) \citealt{Keppler2019}; (3) \citealt{Joyce2023}; (4) \citealt{Muller2018}; (5) \citealt{Mesa2019}; (6)\citealt{Haffert2019};(7) \citealt{Wagner2018}. For additional constraints on the planets' accretion rates, we refer the reader to \citealt{ShibaikeMordasini2024}.} 
\end{deluxetable}

\section{Methods} \label{methods}

\subsection{Disk Parametrization} \label{diskparam}

We carried out planet-disk interaction simulations using the Dusty-FARGO code \citep{Baruteau2019}, which is a 2D hydrodynamic code that solves for the hydrodynamic equations on a radius-azimuth grid. Although the code allows for the placement of dust particles, we do not include dust in this study, as our goal is to assess the orbital architecture of the planets, and not the dust structure. The transport of fluid in the disk is governed by the mass and momentum conservation equations:
\begin{equation}
\frac{\partial \Sigma}{\partial t} + \nabla \cdot (\Sigma \vec{v}) = 0 
\end{equation}
\begin{equation}
\Sigma\left(\frac{\partial \vec{v}}{\partial t} + \nabla\cdot \vec{v}\right)  = - \nabla P - \Sigma \nabla\left(\Phi_{*} + \Phi_{p}\right) + \nabla \Pi
\end{equation}
where $\Sigma$ is the disk surface density, $\vec{v}$ is the vector velocity, $\Phi_{*}$ is the central star's potential, $\Phi_{p}$ is the planet's potential, $P = \Sigma c_s^2$ is the gas pressure and $\Pi$ is the viscous stress tensor. The viscous evolution of the disk requires a viscosity $\nu$ \citep{ShakuraaSunyaev1973}
\begin{equation}
    \nu = \alpha c_{s} H, \label{viscosity}
\end{equation}
where $\alpha$ is a dimensionless parameter, $c_s$ is the isothermal sound speed, and H is the scale height perpendicular to the disk plane at radius R. Here, we adopt best-fit parameter values found by \citet{Keppler2018} and used by \citet{Bae2019} for the PDS 70 disk. The fits were made using radiative transfer models that reproduce the observations in sub-mm/mm wavelengths from ALMA and near-infrared observations from VLT/SPHERE. The simulations use a locally isothermal equation of state, with pre-specified temperatures in every grid cell. This means we do not have to solve an energy equation, and the temperature in the disk is set as a function of radius. The scale height fit gives
\begin{equation}
    \frac{H}{R} = 0.067\left({\frac{R}{22 AU}}\right)^{0.38}. \label{hoverr}
\end{equation}
The disk's integrated temperature is 
\begin{equation}
    T(R) = 44 K (\frac{R}{22 au})^{-0.24}, \label{temperature}
\end{equation}
where R is the distance from the central star, and the local sound speed is
\begin{equation}
    c_s = \sqrt{\frac{k_b T}{\mu m_h}}, \label{cs}
\end{equation}
where $k_b$ is the Boltzmann constant, T is the local temperature, $\mu$ is the mean molecular weight, and $m_h$ is the atomic mass unit of a hydrogen atom. The initial density profile for the disk is parametrized by 
\begin{equation}
    \Sigma_{gas, init}(R) = \Sigma_{c} \left({\frac{R}{R_{c}}}\right)^{-1} \exp\left({-\frac{R}{R_{c}}}\right),
\end{equation}
where $R_c$ is 40 AU and $\Sigma_{c}$ is 2.7 g/$cm^3$ such that the total disk mass is 0.003 $M_\odot$ and the density profile has an exponential tail as a function of radius R. 
At the edge of the radial domain, we adopt a wave-damping zone \citep{deValBorro2006} to suppress reflection of waves at the boundaries. 
The simulation is run on a grid that spans 2.2 to 198 AU in the radial direction and the full 2$\pi$ range in the azimuthal direction in order to cover the extent of the observed disk. The resolution is 936 grid spaces in the azimuthal direction and 672 grid spaces in the radial direction, following \citealt{Bae2019}. Dusty-FARGO's N-body integrator is a Runge Kutta fifth order scheme. The timestep is governed by the CFL condition, as described in \citealt{BenitezLlamblay2016}.

\subsection{Photoevaporation Prescription} \label{photoevap}
 \begin{figure*}
  \begin{center}
\centerline{\includegraphics[width=\textwidth]{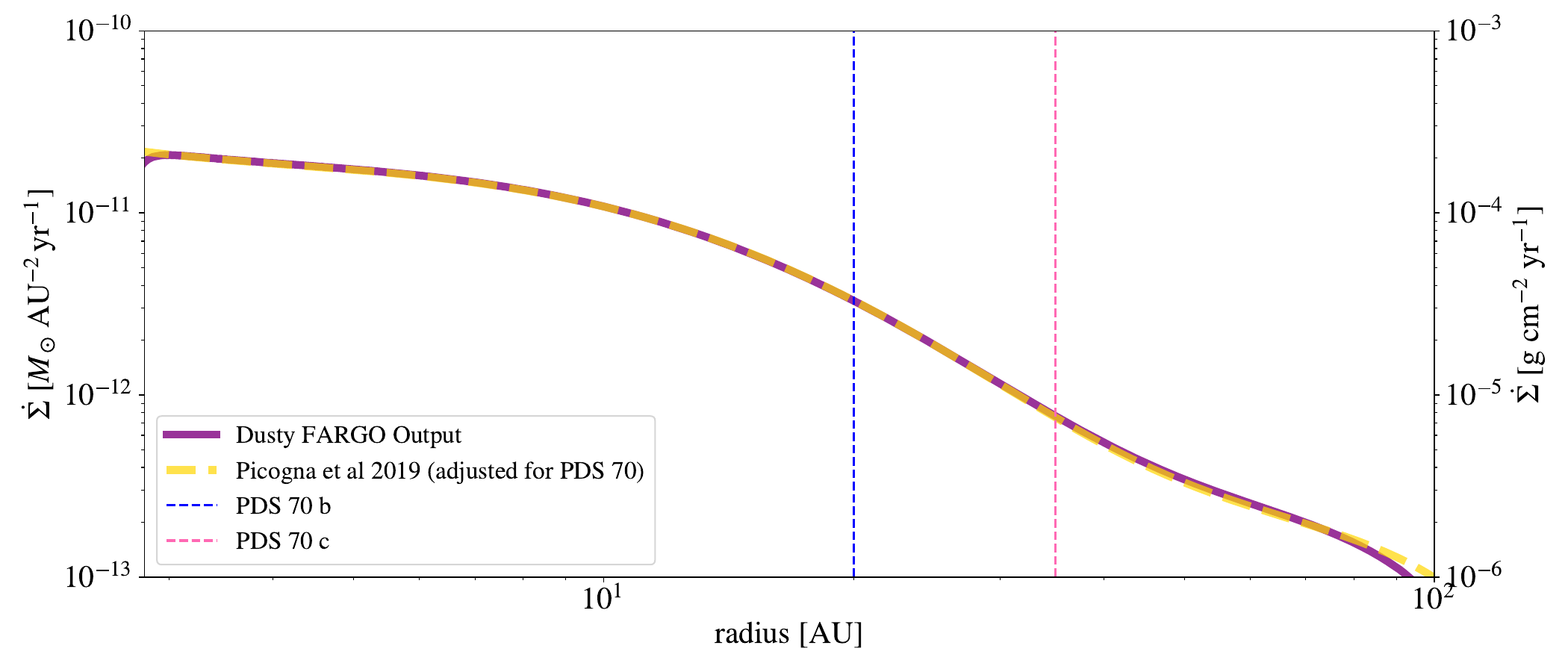}}
\caption{Comparison of the input photoevaporation prescription presented in \citealt{Picogna2019} (adjusted for PDS 70's average X-ray luminosity value) and the output obtained in Dusty-FARGO after implementation. This step serves as a verification that the PDS 70 disk is losing mass according to the prescription. Initial locations of PDS 70 b and c are marked in blue and pink respectively. }
\label{fig:photoevaporationpng}
  \end{center}
\end{figure*}

\par
 We follow the photoevaporation prescription presented in \cite{Picogna2019} (in particular, the radial profile of density loss presented in their equations 2, 3 and 5). Their prescription uses radiative transfer modeling of low-mass stars ($\sim$ 0.7 \Msol) with small disk mass ($<$ 1\% of stellar mass), making it well suited for the PDS 70 case. Here, the prescription is implemented for the case of PDS 70, where the central star has an average X-ray luminosity of $1.37 \times 10^{30}$ ergs/s \citep{Joyce2023}. The photoevaporation is implemented as a source term in the Dusty-FARGO code. At each timestep, the photoevaporation removes a certain amount of the surface density from the disk, varying radially according to the implemented prescription from \citealt{Picogna2019}.\par
We first verify that the disk's photoevaporation is correctly implemented by simulating the system without the PDS 70 b and c planets and disk viscosity. Figure \ref{fig:photoevaporationpng} shows the photoevaporation prescription input compared to Dusty-FARGO's output after implementation. We obtain the output from Dusty-FARGO by subtracting the initial density from the final density at a given time step. For the time step shown in Figure \ref{fig:photoevaporationpng}, we chose the final step to be about 500 years of integration, so near the beginning of the simulation, well before the disk density reaches a density floor. The boundaries present a smaller photoevaporation rate than the prescription due to the application of a wave damping zone at the boundaries in order to supress wave reflection \citep{deValBorro2006}. We set a density floor value for the simulation such that the disk does not exhibit negative density as the photoevaporation progresses. We set the density floor to be  $<$$1 \times 10^{-5} g$ $cm^{-2}$ \citep{Portilla-Revelo2022}.  
 \par
We include a photoevaporation ``efficiency", $\kappa$, which varies the photoevaporation rate by a factor of 0, 0.1, 1 and 10. This is motivated by the highly uncertain and variable (on the order of days) X-ray luminosity of T Tauri stars. This variation was reported for PDS 70 (luminosity increased up to a factor of $\sim$ 3 due to a flare, and varied by a range of two orders of magnitude) \citep{Joyce2023}, and for other young stars \citet{Feigelson2002, Caramazza2007}. \par
\subsection{Initial Conditions and System Evolution} \label{planetinit}

We add the PDS 70 b and c planets to the simulation and allow them to dynamically interact with each other and the disk.  We use an initial semi-major axis of 20 AU for PDS 70 b and 35 AU for PDS 70 c following \citet{Bae2019}, which is consistent with most orbital solutions and the projected separations of both planets. Although this is our chosen starting location, changing these values to any separations currently consistent with observations is not expected to change the outcome of the orbital architectures. These chosen separations were also capable of reproducing the observed dust structure in the disk \citep{Bae2019}. Furthermore, previous studies that explored a broader range of initial separations found similar outcomes when evolving the planets from their possible formation locations to their current orbits \citep{Toci2020}.\par
We fix the mass of PDS 70 b to 7 $M_{Jup}$, consistent with observational values, which range from 4--17 $M_{Jup}$ \citep{Muller2018,Mesa2019, Haffert2019, Mesa2019}. We test three values for PDS 70 c's mass, setting it to 4, 7 and 10 $M_{Jup}$, all consistent with observations (4--12 $M_{Jup}$; \citealt{Haffert2019, Wang2021, Mesa2019}). Given the large uncertainties on PDS 70's planet masses, we choose these values with the goal of testing cases where the outer planet is, respectively, less massive, equally massive and more massive than the inner planet, as the mass ratio ($M_{outer}/M_{inner}$) affects the final migration direction \citep{Masset2001, Snellgrove2001}.
\par
Since the $\alpha$ parameter used can also change the direction of planetary migration, we test different $\alpha$ values in order to assess how it affects the final evolution and stability of the system. We test three values of $\alpha$ in our simulations for one of the mass cases (0, $10^{-3}$ and $10^{-2}$) in order to examine viscous transport timescales at several orders of magnitude. The nominal $\alpha$ parameter is set to $10^{-3}$, as was found in \citep{Bae2019} to reproduce the dust structure in the disk, and consistent with \citealt{Portilla-Revelo2023}'s findings using ALMA observations regarding the gap depth ($\alpha < 5\times10^{-3}$). We keep $\alpha$ independent of the radius for our analysis, such that it is uniform throughout the disk. \par
Therefore, we have 20 total combinations of parameters for our PDS 70 simulations. The first three columns of Table \ref{tbl:initial_conds} summarize the suite of initial conditions explored in our simulations. 
\par

\begin{deluxetable*}{ccccccccc} 
\tablecaption{Initial and Final Conditions for PDS 70} 
\label{tbl:initial_conds}
\tablewidth{20pt}
\tablecolumns{9}
\tabletypesize{\scriptsize}
\tablehead{\colhead{ $\kappa$ } & \colhead{$M_c$ ($M_{Jup})$} & \colhead{$\alpha$} & \colhead{$a_b$ (AU)} & \colhead{$a_c$ (AU)} & \colhead{$e_b$} & \colhead{$e_c$} & \colhead{$P_c/P_b$} & \colhead{$\theta_{amp}$}}
\startdata
0 &  4 & $10^{-3}$ & $21.37 \pm 0.01$ & $35.08 \pm 0.55$ & $0.08 \pm 0.003$ & $0.02 \pm 0.006$ & $2.10 \pm 0.03$ & $11.79$ \\
$\frac{1}{10}$ &  4 & $10^{-3}$ & $20.10 \pm 0.02$ & $33.33 \pm 0.45$ & $0.06 \pm 0.001$ & $0.02 \pm 0.005$ & $2.14 \pm 0.02$ & $10.33$ \\
1 &  4 & $10^{-3}$ &  $20.22 \pm 0.02$ & $33.23 \pm 0.45$ & $0.07 \pm 0.007$ & $0.02 \pm 0.005$ & $2.11 \pm 0.02$ & $20.66$ \\
10 &  4 & $10^{-3}$& $19.79 \pm 0.02$ & $33.94 \pm 0.39$ & $0.03 \pm 0.015$ & $0.03 \pm 0.006$ & $2.25 \pm 0.02$ & $70.63$ \\
0 &  7& $10^{-3}$ & $21.29 \pm 0.03$ & $34.99 \pm 0.54$ & $0.12 \pm 0.003$ & $0.02 \pm 0.004$ & $2.11 \pm 0.03$ & $7.48$ \\
$\frac{1}{10}$ &  7 & $10^{-3}$ & $19.42 \pm 0.03$ & $32.18 \pm 0.39$ & $0.09 \pm 0.008$ & $0.02 \pm 0.004$ & $2.14 \pm 0.02$ & $18.89$ \\
1 &  7 & $10^{-3}$ & $20.03 \pm 0.05$ & $32.92 \pm 0.44$ & $0.12 \pm 0.01$ & $0.03 \pm 0.01$ & $2.11 \pm 0.03$ & $25.73$ \\
10 &  7 & $10^{-3}$ & $19.66 \pm 0.04$ & $33.72 \pm 0.38$ & $0.05 \pm 0.02$ & $0.03 \pm 0.006$ & $2.25 \pm 0.02$ & $58.37$ \\
0 &  10 & $10^{-3}$ & $18.88 \pm 0.06$ & $31.06 \pm 0.36$ & $0.15 \pm 0.02$ & $0.07 \pm 0.02$ & $2.11 \pm 0.02$ & $35.81$ \\
$\frac{1}{10}$ &  10 & $10^{-3}$ & $19.19 \pm 0.09$ & $31.84 \pm 0.38$ & $0.12 \pm 0.02$ & $0.02 \pm 0.004$ & $2.14 \pm 0.03$ & $30.06$ \\
1 &  10 & $10^{-3}$ & $19.70 \pm 0.09$ & $32.94 \pm 0.41$ & $0.10 \pm 0.02$ & $0.03 \pm 0.009$ & $2.16 \pm 0.03$ & $42.55$ \\
10 &  10 & $10^{-3}$ & $19.47 \pm 0.07$ & $33.76 \pm 0.37$ & $0.07 \pm 0.03$ & $0.03 \pm 0.008$ & $2.29 \pm 0.03$ & $71.47$ \\
0 &  4 & $10^{-2}$ & $25.99 \pm 0.28$ & $42.43 \pm 0.89$ & $0.10 \pm 0.002$ & $0.03 \pm 0.009$ & $2.09 \pm 0.05$ & $11.23$ \\
$\frac{1}{10}$ &  4 & $10^{-2}$ & $24.11 \pm 0.21$ & $39.48 \pm 0.75$ & $0.09 \pm 0.004$ & $0.03 \pm 0.01$ & $2.09 \pm 0.04$ & $16.48$ \\
1 &  4 & $10^{-2}$ & $20.67 \pm 0.02$ & $33.76 \pm 0.49$ & $0.09 \pm 0.006$ & $0.04 \pm 0.01$ & $2.09 \pm 0.03$ & $24.09$ \\
10 &  4 & $10^{-2}$ & $19.94 \pm 0.02$ & $33.78 \pm 0.41$ & $0.04 \pm 0.01$ & $0.03 \pm 0.006$ & $2.21 \pm 0.02$ & $40.79$  \\
0 &  4 & 0 & $20.83 \pm 0.01$ & $34.42 \pm 0.51$ & $0.06 \pm 0.002$ & $0.02 \pm 0.005$ & $2.13 \pm 0.03$ & $12.99$ \\
$\frac{1}{10}$ &  4 & 0 & $20.28 \pm 0.03$ & $34.20 \pm 0.43$ & $0.04 \pm 0.02$ & $0.02 \pm 0.005$ & $2.19 \pm 0.02$ & $50.93$ \\
1 &  4 & 0 & $20.01 \pm 0.02$ & $33.20 \pm 0.42$ & $0.06 \pm 0.01$ & $0.02 \pm 0.005$ & $2.14 \pm 0.02$ & $26.10$ \\
10 &  4 & 0 & $19.73 \pm 0.01$ & $33.92 \pm 0.39$ & $0.03 \pm 0.008$ & $0.03 \pm 0.006$ & $2.26 \pm 0.02$ & $52.39$ \\
\enddata
\tablecomments{We keep $M_b$ fixed to 7 $M_{Jup}$ in every configuration. The initial locations of the planets are 20 and 35 AU for b and c, respectively. The reported final values are computed as the median and 68th percentile uncertainties from the last 10\% of the duration of the hydrodynamic simulations. The astrocentric elements are natural as the raw output from FARGO, but the reader should be aware that they are more time-variable than, e.g., the Jacobi coordinates  \citep{Zurlo2022}. We quote them to show and compare the solutions derived for different parameter sets. The use of astrocentic coordinates is a way to encode Cartesian coordinates/momentums (velocities). $\theta_{amp}$ denotes the average oscillation amplitude of the 2:1 resonant angle in degrees. }
\end{deluxetable*}

\subsection{When the disk is gone: N-body Simulations}
After the protoplanetary disk is sufficiently photoevaporated (i.e., the density floor extends beyond the planets' Hill radii), we aim to analyze how the planetary system will evolve over time using N-body simulations. Since our goal is to explore long-term N-body integrations, we require higher accuracy than the built-in Runge-Kutta scheme implemented in FARGO, which is mostly useful for shorter integration times (i.e. the photoevaporative lifetime of the disk). We therefore use the N-body code REBOUND \citep{Rein_2015}. We use the fast, non-symplectic, 15th-order IAS15 integrator \citep{Rein2015} to integrate the system and assess stability over billion-year timescales. IAS15 utilizes an adaptive timestep with error control based on local truncation estimates, enabling high precision while maintaining computational efficiency. The integrator is based on a Gauss-Radau collocation scheme, which allows for accurate handling of close encounters and long-term integration with machine precision \citep{Rein2015}. The reason for probing such long timescales is to make predictions on whether older systems are likely to host multiple ultra-massive, widely separated gas giants in wide orbits. 
\par
Since the initial disk mass is $\sim$ 3 $M_{Jup}$, and it dissipates with time to 0 -- 2.4 $M_{Jup}$ during our FARGO integrations, planet-planet interactions are expected to dominate the evolution process. Our main goal with running the hydrodynamic simulations is to evolve the system from a set of orbital parameters where the planets can migrate into the 2:1 mean-motion resonance rather than drawing parameters from currently unconstrained orbit fits. For that reason, we only consider the planet-planet and planet-star interactions in our N-body simulations with REBOUND.
\par
We use the chaos indicator tool, MEGNO (or Mean Exponential Growth \added{Factor} of Nearby Orbits) \citep{Cincotta2003}, to compute whether the orbital parameters of a planetary system remain stable over time. If a system is chaotic, in the sense of non-zero Maximal Lyapunov Exponent, two initial configurations that start near each other will have exponentially diverging trajectories. The MEGNO value in REBOUND is calculated by placing a shadow particle with slightly perturbed initial conditions and considering the displacement vector of the two particles (defined as \textbf{$\delta_i$}) and obtaining the equations of motion using the variational principle on the trajectories. If the MEGNO value goes to infinity, the system is unstable. If it converges to $\leq$2, the system is stable.
\par
Here we consider stable configurations to have MEGNO values that converge between $1.95 \leq Y \leq 2.05$ (e.g. \citealt{Gozdziewski2014}). Since MEGNO is a fast chaos indicator, it is capable of assessing chaotic trajectories in shorter timescales without requiring direct long-term integration. For that reason, we only integrate and track orbital parameters for the system for 100 Myr, which is sufficient to identify chaos for 10--100$\times$ longer timescales such as 1 -- 10 Gyr \citep{Gozdziewski2001}.
In order to assess the system's stability in a statistical manner (in particular since the planets' osculating elements can vary significantly in Dusty-FARGO's N-body scheme), we randomly draw 100 outputs from the last 10\% of integration in FARGO for each configuration after the disk is dissipated around the planets (e.g. after $\sim$ 22,000 years for $\kappa = 10$ and $\sim$ 0.22 Myr for $\kappa = 1$, etc), and assess their MEGNO parameter to obtain a distribution of ``stability likelihood" for each configuration. We test ``two'' (the first one contains 20 within itself) different configurations:

\begin{enumerate}
    \item The two-planet configurations from Dusty-FARGO's outputs (20 total when considering $\kappa$, $M_c$ and $\alpha$ variations)
    \item The two-planet configurations from orbital fits, where orbital parameters are drawn from priors given astrometric positions/radial velocities 
\end{enumerate}

We choose these models in order to compare the stability expectations from dynamical models of the protoplanets from the disk case with the observational results from orbit fits that use pure N-body integrations and do not consider planet-disk interactions. 

\section{Results} \label{results}

\subsection{$M_c$ and $\kappa$ Dependence} \label{planetparam-mckappa}
\begin{figure*}
    \centering
    \centering{{\includegraphics[width= \textwidth]{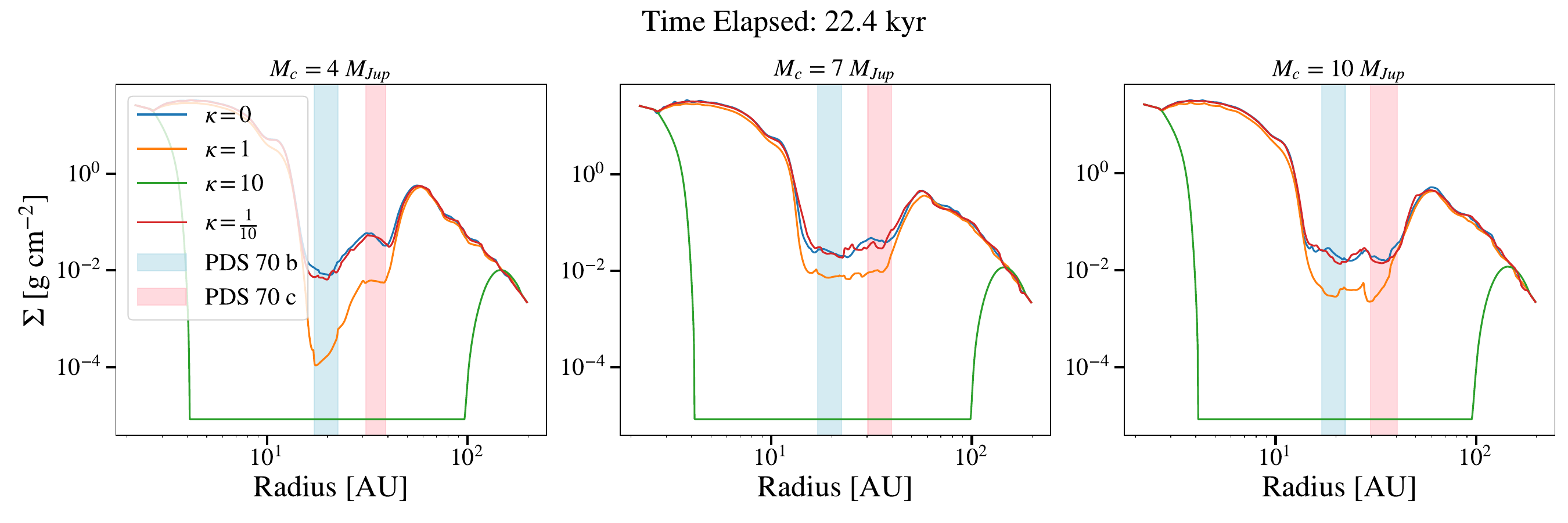} }}%
    \qquad
    \centering{{\includegraphics[width= \textwidth]{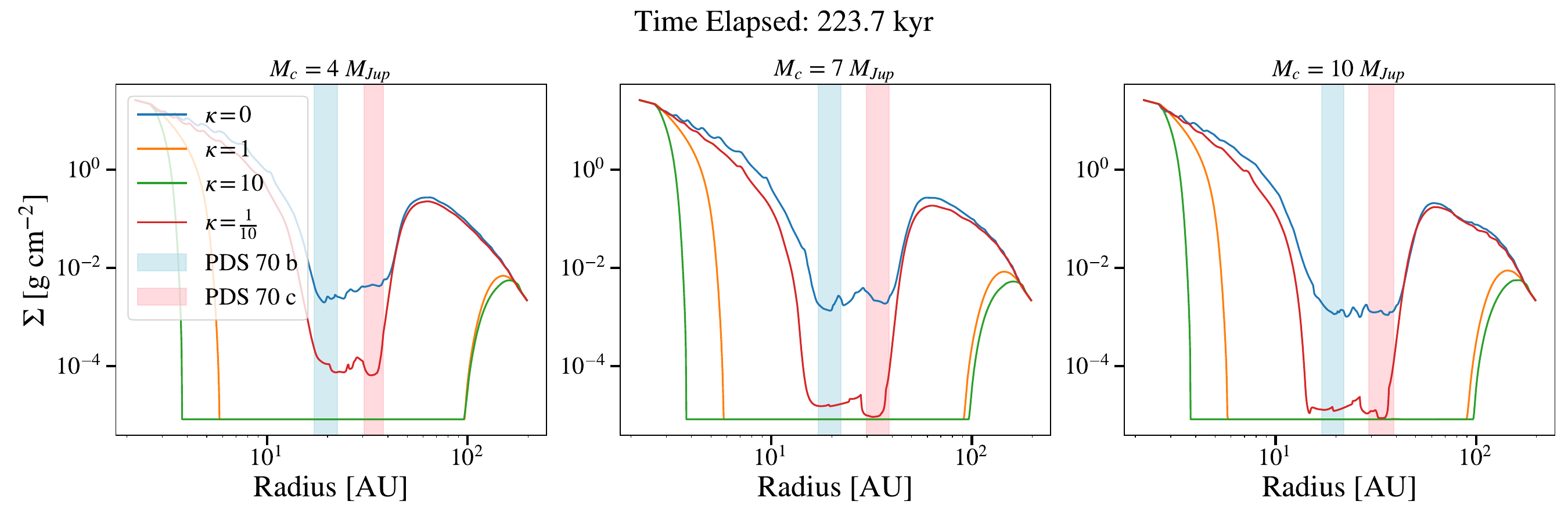} }}%
    \qquad
    \centering {{\includegraphics[width=\textwidth]{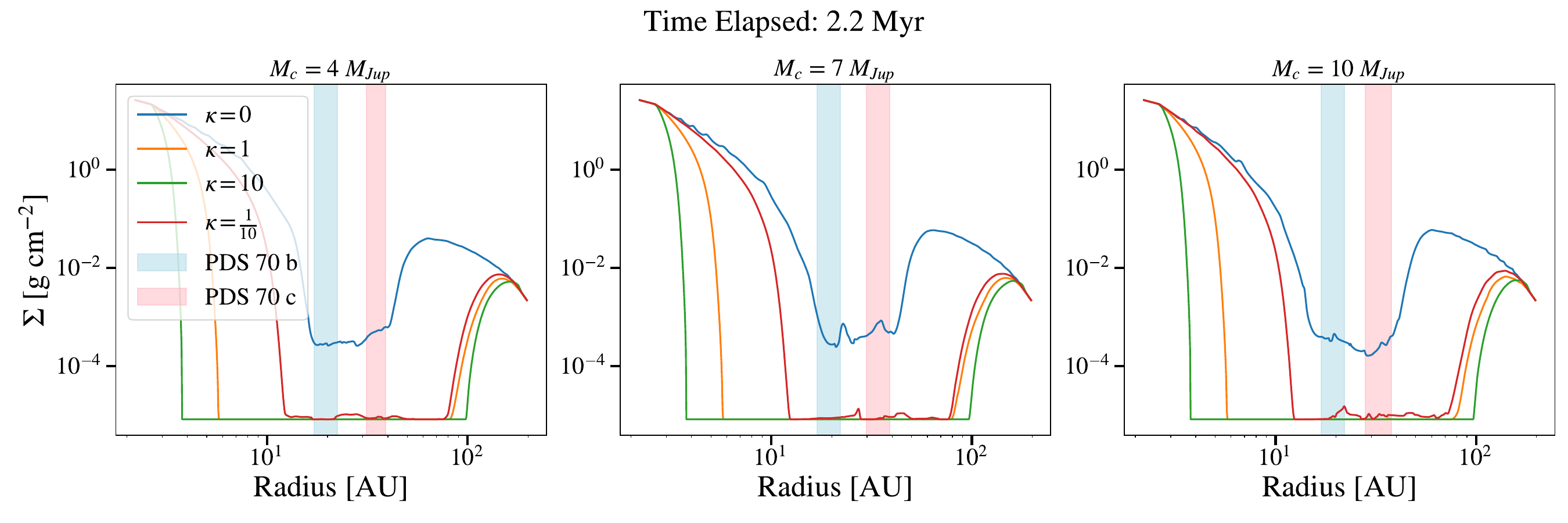} }}
    \caption{The average density of the PDS 70 disk as a function of radius for 22 kyr (top panels), 0.22 Myr (middle panels) and 2.2 Myr (bottom panels) integration times. The Hill radius extension of the two planets is shown as the light blue and light pink shaded regions. The disk profile dissipation at different rates is represented in different colors: blue ($\kappa$ = 0), red ($\kappa$ = $\frac{1}{10}$), orange ($\kappa$ = 1) and green ($\kappa$ = 10). }%
    \label{fig:densplots}
\end{figure*}

In order to verify that the disk density is mostly dissipated around the planets by the end of the simulations, we plot the surface density as a function of distance from the central star throughout time steps, shown in Figure \ref{fig:densplots}. Using the Hill radius
\begin{equation}
    R_H = a (1 - e) \left(\frac{m_p}{3M_s}\right)^{1/3},
\end{equation}
where $m_p$ is the mass of the planet, $a$ and $e$ are the semi-major axis and eccentricity of the planet, and $M_s$ is the mass of the star, we mask a region that corresponds to the Hill ``circle" (since these simulations are two-dimensional) around each planet.  We find that the disk is dissipated from the planets' region for all values of $\kappa$ within $\sim$ 2.2 Myr. \par
\begin{figure*}[ht!]
  \begin{center}
\centerline{\includegraphics[width=\textwidth]{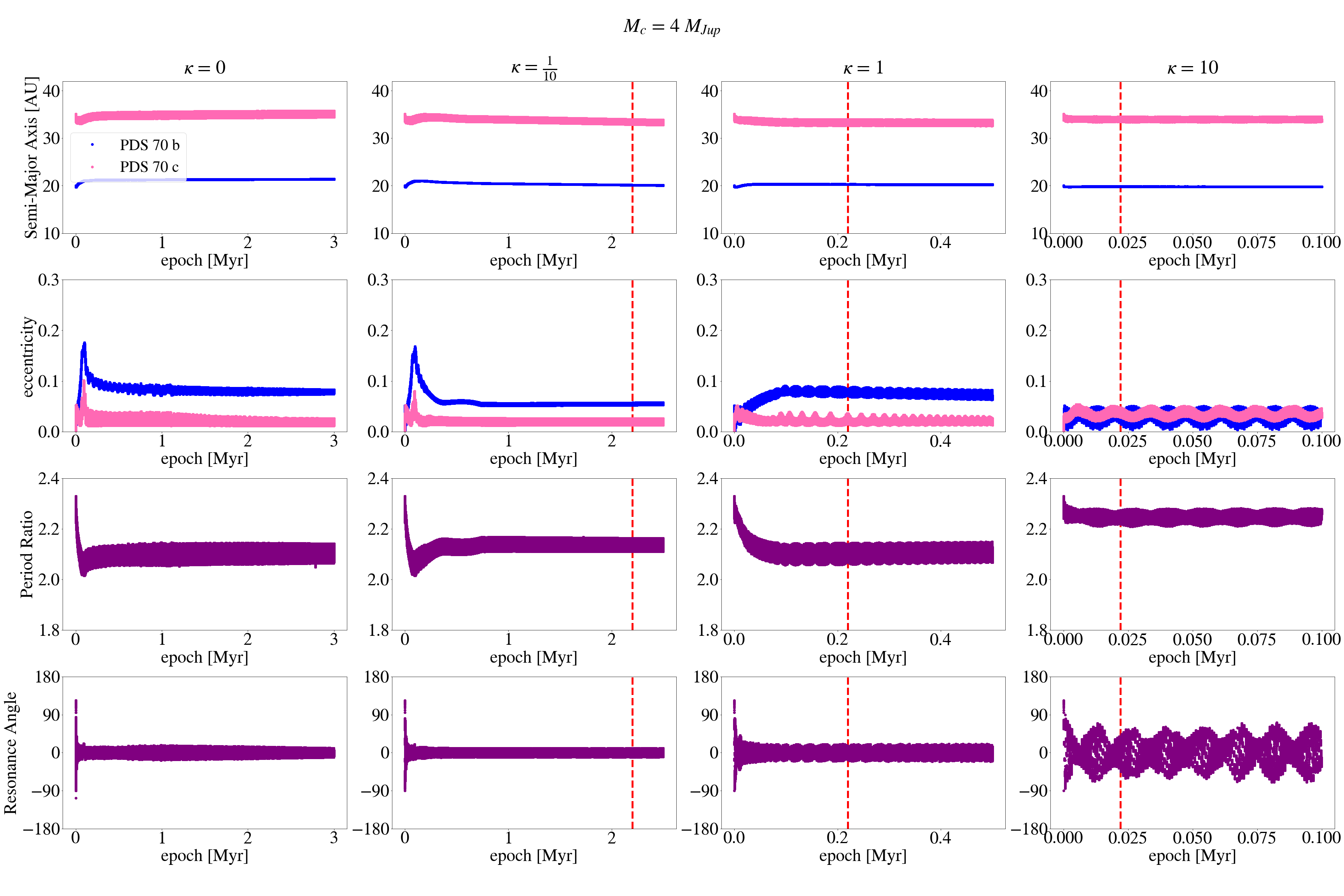}}
\caption{Evolution of the planets' b (blue) and c (pink) orbits under different photoevaporation rates $\kappa$ for the protoplanetary disk. Here, $M_c$ is set to 4 $M_{Jup}$. We plot the semi-major axis, eccentricity, period ratio and resonant angle $ \theta_{c:b}$. The vertical dotted lines in each panel correspond to timescales for the disk's full dissipation around the planets' location. For examples of these surface densities around these timescales, please refer to Figure \ref{fig:densplots}. }
\label{fig:c4}
  \end{center}
  \end{figure*}
\begin{figure*}[ht!]
  \begin{center}
\centerline{\includegraphics[width=\textwidth]{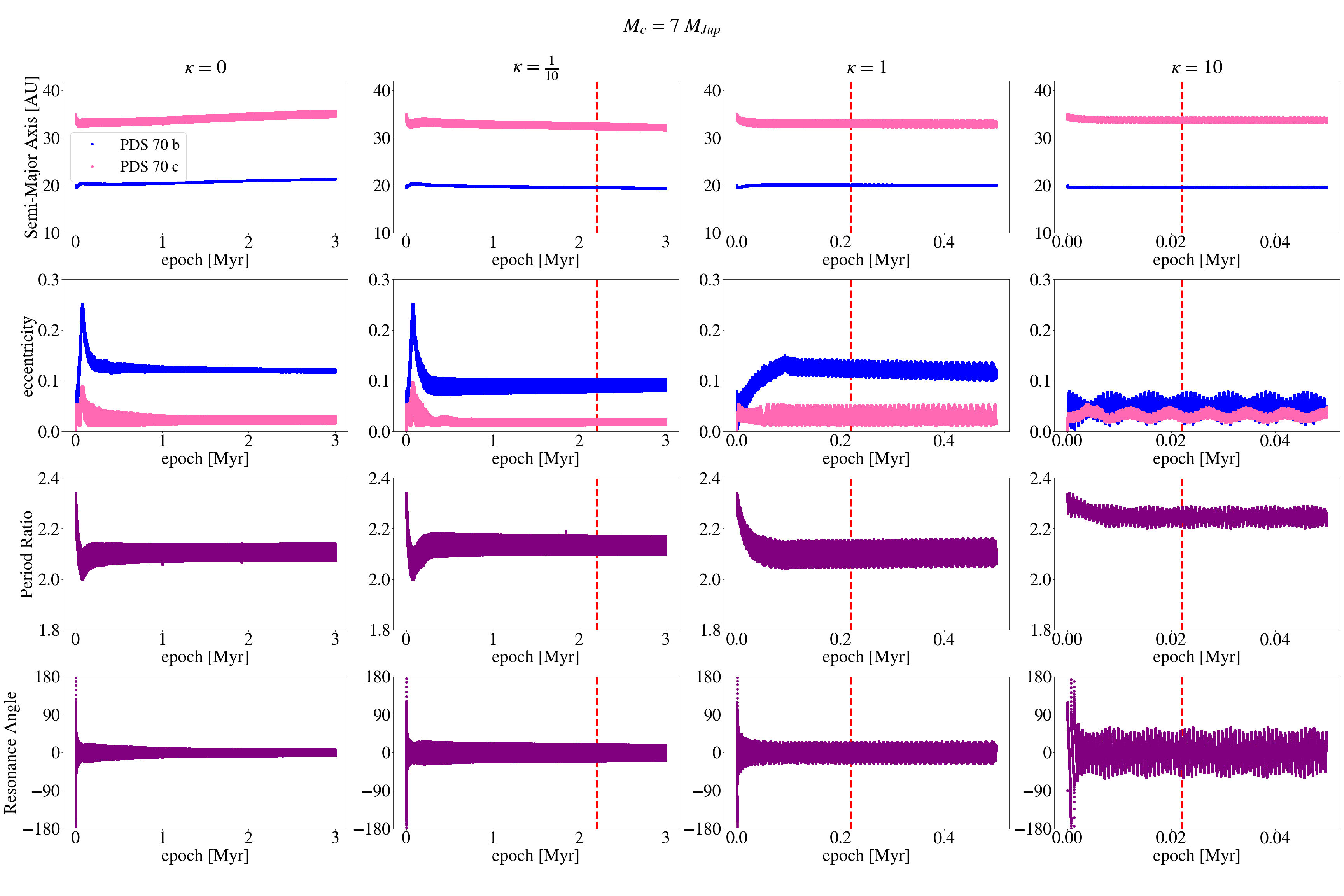}}
\caption{Same as figure \ref{fig:c4}, but with $M_c$ set to 7 $M_{Jup}$.}
\label{fig:c7}
  \end{center}
  \end{figure*}
\begin{figure*}[ht!]
  \begin{center}
\centerline{\includegraphics[width=\textwidth]{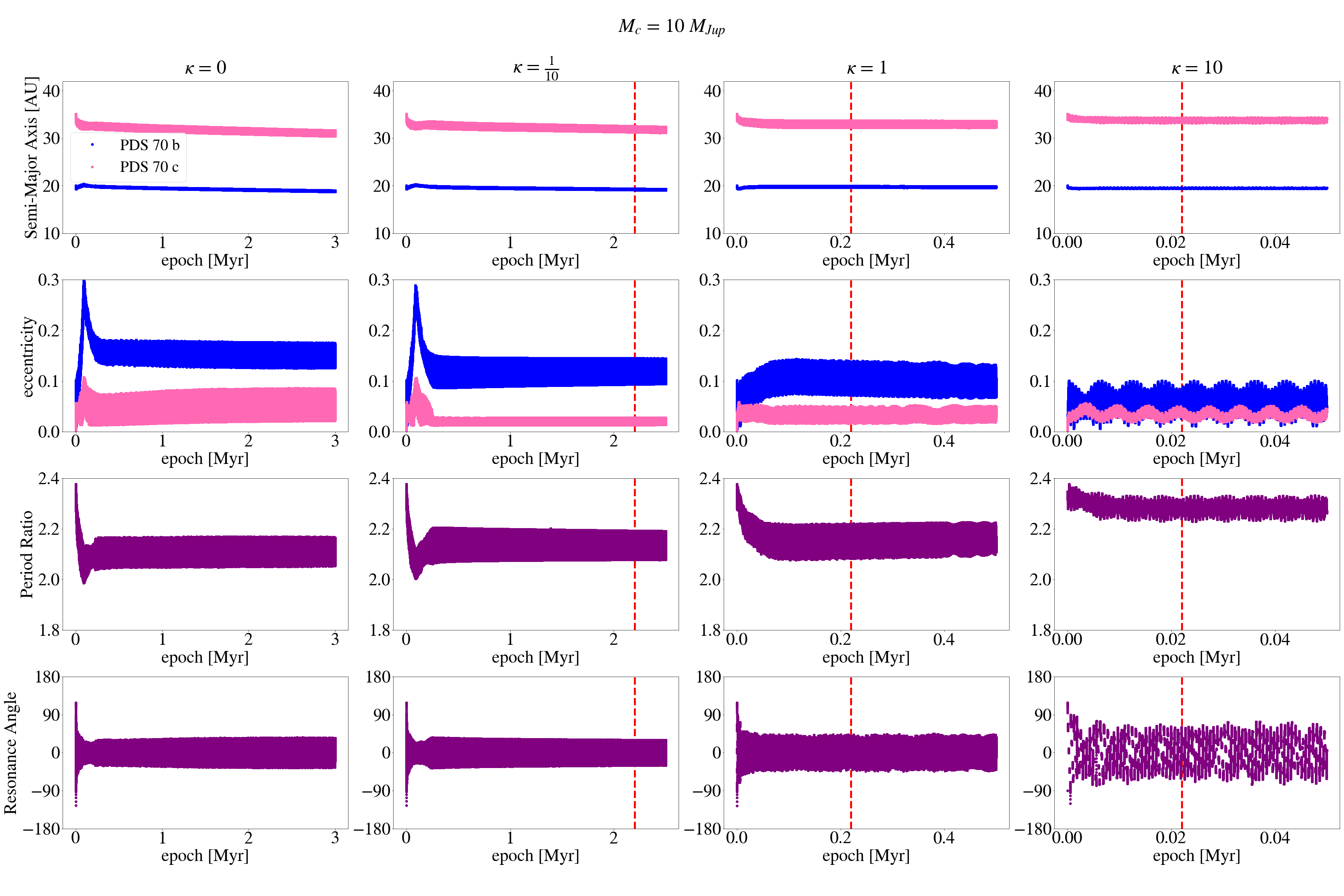}}
\caption{Same as figure \ref{fig:c4}, but with $M_c$ set to 10 $M_{Jup}$.}
\label{fig:c10}
  \end{center}
  \end{figure*}

We track 4 orbital parameters for the planets: the semi-major axis, eccentricity, period ratio ($\frac{P_c}{P_b}$) and resonance angle ($\theta_{c:b}$). Our final orbital parameter results for each case are presented in the final columns of Table \ref{tbl:initial_conds}. The initial placement of the planets in the disk is close to a 2:1 mean-motion resonance (the true ratio is slightly larger). Having a near-integer period ratio does not signify that the planets are necessarily in resonance. Mean motion resonance is defined by the resonant angle librating over a value that is dependent on the specific resonance. Here, one of the resonant angles for a 2:1 resonance between planet PDS 70 c and planet PDS 70 b is \citep{Bae2019, Wang2021}:

\begin{equation} \label{eq:7}
    \theta_{c:b} = \lambda_b - 2\lambda_c +  \varpi_b,
\end{equation}

where the longitude of periastron is $\varpi = \Omega + \omega$ and $\lambda = \varpi + M $ is the mean longitude, defined using $M$, the mean anomaly and $\Omega$, the longitude of the ascending node. The other angle would simply be $\theta_{b:c}$, using the values relative to the outer planet's argument of periastron. Our results from varying the mass of the outer planet, $M_c$, and the photoevaporation rate, $\kappa$, are shown in Figures \ref{fig:c4}, \ref{fig:c7} and \ref{fig:c10}. 
\par

\begin{figure*}[ht!]
  \begin{center}
\centerline{\includegraphics[width=\textwidth]{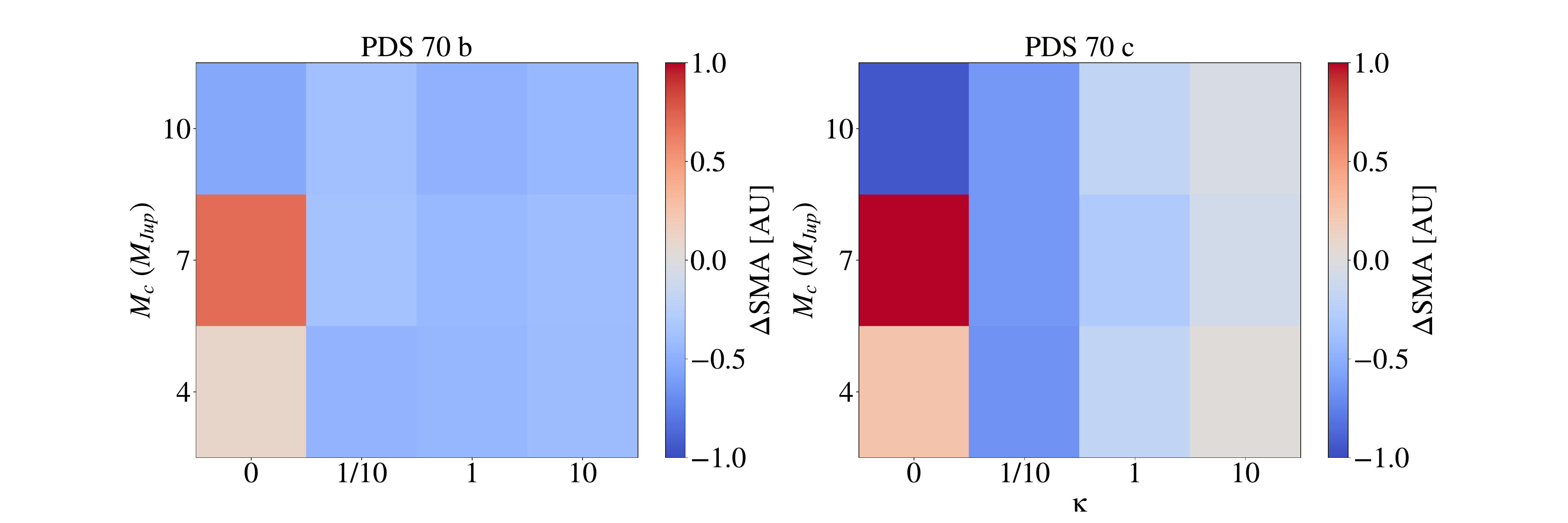}}
\caption{Final migration contour of the inner planet (left panel) and outer planet (right panel), as a function of photoevaporation rate $\kappa$ (x-axis) and mass of the outer planet $M_c$ (y-axis).}
\label{fig:sma-mass}
  \end{center}
  \end{figure*}

The migration direction of the planets is also dependent on the mass ratio of b and c, along with the $\kappa$ value. This is illustrated in Figure \ref{fig:sma-mass}. The $\Delta SMA$ is calculated as the final average $SMA_f - SMA_{500,000}$, where $SMA_{500,000}$ is the semi-major axis at 500,000 years of integration and $SMA_f$ is the final semi-major axis at 2 Myr of integration. In the case where the disk does not dissipate at all ($\kappa$ = 0), both planets migrate outwards if the inner planet is equally massive or more massive than the outer planet ($M_c$ is 4 or 7 $M_{Jup}$), and inwards if the outer planet is more massive (10 $M_{Jup}$). The planets get locked into resonance in all cases, although their resonant angle libration amplitude increases with increasing $M_c$ mass. \par
In the case where the disk dissipates rapidly  ($\kappa$ = $\frac{1}{10}$, 1 or 10), the planets migrate inwards in all cases. The strength of their migration is directly proportional to the $\kappa$ value: as $\kappa$ increases, the migration rate decreases. Resonance locking occurs in all cases, with increasing amplitude as $\kappa$ increases. \par
We also evaluate the eccentricities of the two planets as a function of $M_c$ and $\kappa$. In the case where the disk does not dissipate at all ($\kappa$ = 0), the eccentricity of the inner planet gets excited. The value it gets excited to ($\approx$ 0.18, 0.25 and 0.30) is directly proportional to the outer planet's mass, with higher outer planet masses providing higher values of eccentricity excitation. This is a consistent result with \citealt{Bae2019}'s findings, and is a characteristic of 2:1 resonance locking. When the outer planet is more massive than the inner planet, its eccentricity also gets slightly excited, although to smaller values ($\approx$ 0.1 initially). However, in all cases the eccentricity of the inner planet gets dampened after this initial excitation, to about 0.1, 0.12 and 0.15 respectively.\par
In the case where the disk dissipates within $\approx$ hundreds of thousands of years or a few Myr ($\kappa$ = 1/10 or $\kappa = 1$), a similar phenomenon occurs, but with lower eccentricity excitations for the inner planet. If $\kappa = 10$, the eccentricities of the inner planet do not go above 0.08.
\par
Notably, when the disk dissipates very slowly or not at all ($\kappa = \frac{1}{10}$ and $\kappa = 0$), the eccentricities of the planets stay the same after about 0.5 Myr. In the cases where $\kappa = 1$ and $\kappa = 10$, the eccentricities of the inner and outer planet get dampened to $\leq$ 0.05 after the initial excitation.

\subsection{Viscosity Variations: The $\alpha$ value} \label{alpha-results}
\begin{figure*}[ht!]
  \begin{center}
\centerline{\includegraphics[width=\textwidth]{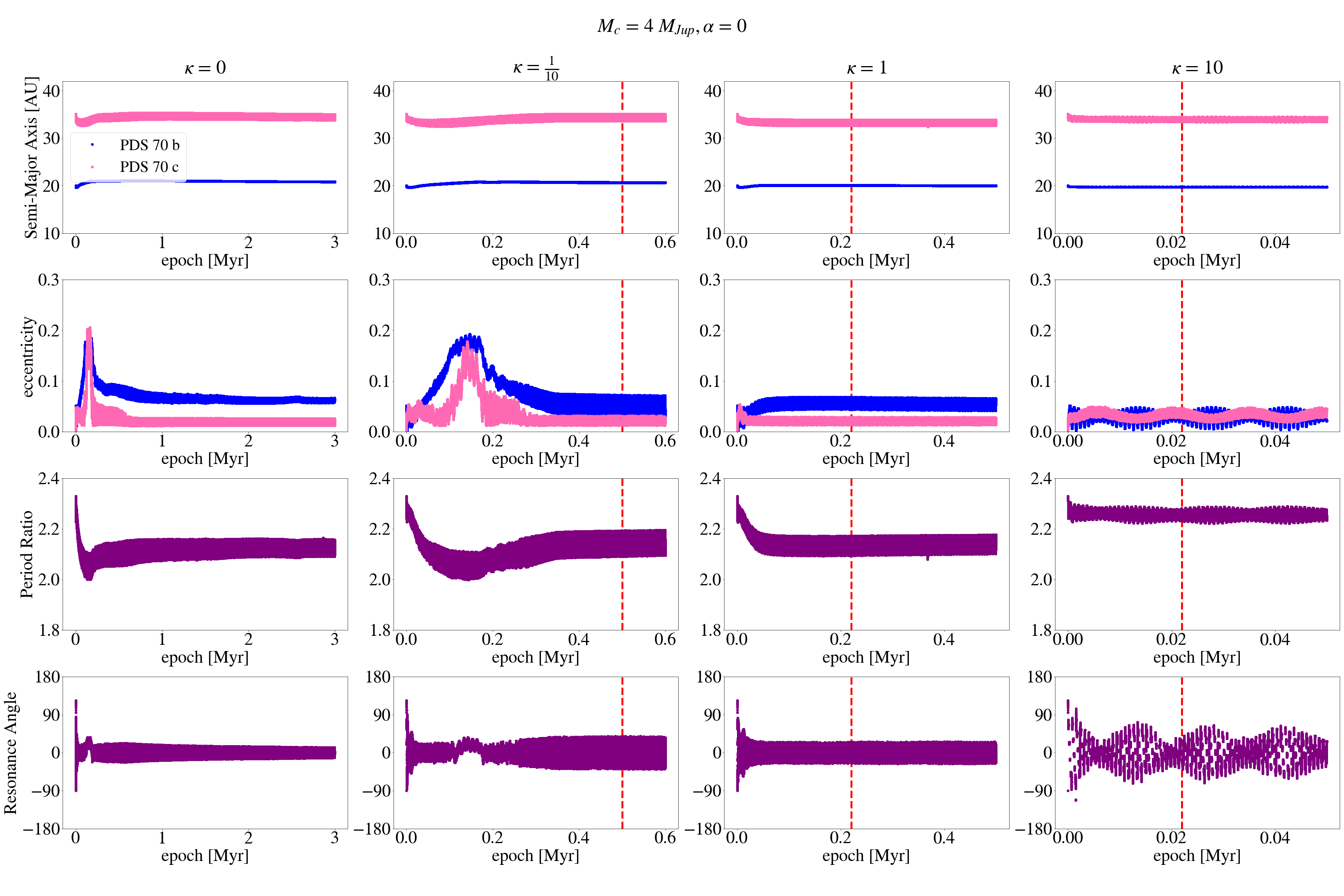}}
\caption{Evolution of the planets' b and c orbits over 5 Myr ($\kappa$ = 1, $\frac{1}{10}$) and 2 Myr ($\kappa$ = 1, 10) under different photoevaporation rates $\kappa$ for the protoplanetary disk. Here, $M_c$ is set to 4 $M_{Jup}$ and $\alpha$ is set to 0.}
\label{fig:alpha0}
  \end{center}
  \end{figure*}

\begin{figure*}[ht!]
  \begin{center}
\centerline{\includegraphics[width=\textwidth]{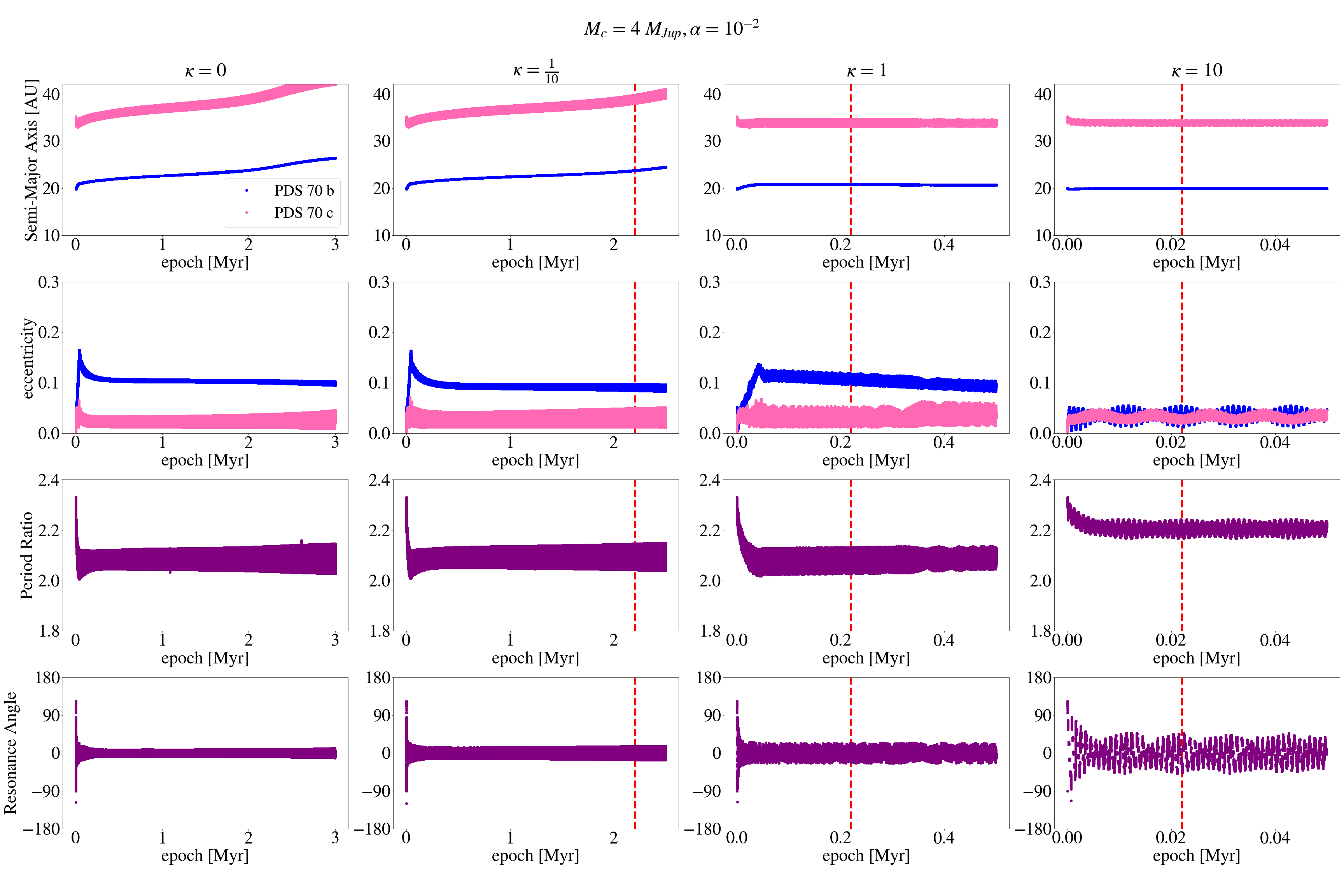}}
\caption{Same as Figure \ref{fig:alpha0}, but with $\alpha$ set to $10^{-2}$.}
\label{fig:alpha1e-2}
  \end{center}
  \end{figure*}

For the case where $M_c$ is 4 $M_{Jup}$, we vary the $\alpha$ viscosity parameter. We include $\alpha$ values of $10^{-3}$, 0, and $10^{-2}$, such that the viscous timescale of the disk varies over several orders of magnitude. Our results can be found in Figures \ref{fig:c4}, \ref{fig:alpha0}, and \ref{fig:alpha1e-2}. \par
   \begin{figure*}[ht!]
  \begin{center}
\centerline{\includegraphics[width=\textwidth]{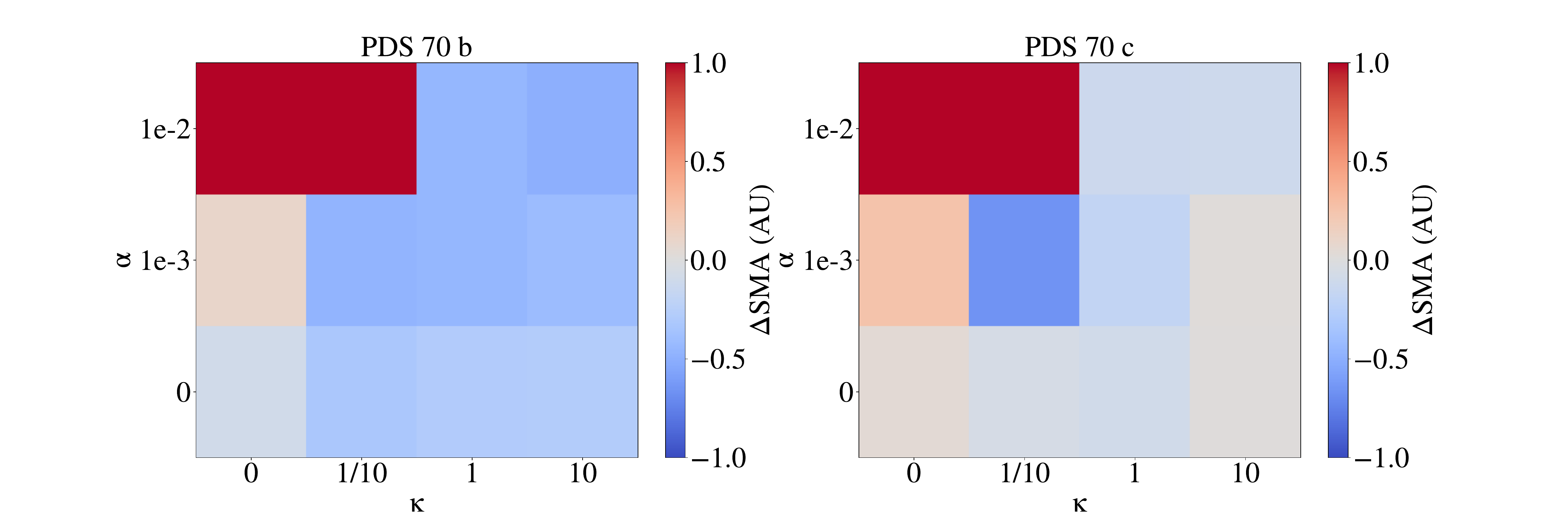}}
\caption{Contour plot of the migration of PDS 70 b (left) and c (right) as a function of disk photoevaporation rate $\kappa$ and viscosity parameter $\alpha$. Red colors represent outward migration while blue colors represent inward migration. Very small migration values, near 0, appear gray. In this simulation, the mass of planet c was set to 4 $M_{Jup}$.}
\label{fig:alphacontour}
  \end{center}
  \end{figure*}

From equations \ref{viscosity} -- \ref{cs}, the viscous timescale is given by
\begin{equation}
    t_{viscous} = \frac{R^2}{\nu}.
\end{equation}
\par

For $\alpha$ of $10^{-3}$, the viscous timescale is in the order of a few Myr (specific values depend on the radial location in the disk). For an order of magnitude larger $\alpha$, it is a few hundreds of thousands of years. For $\alpha$ of 0, there is no viscosity, and the disk is essentially incapable of refilling the gap formed by the two gas giants.
\par
\begin{deluxetable*}{ccccc} 
\tablecaption{Timescales for the PDS 70 System} 
\label{tbl:timescales}
\tablewidth{20pt}
\tablecolumns{5}
\tabletypesize{\scriptsize}
\tablehead{\colhead{ $\kappa$ } & \colhead{$M_c$ ($M_{Jup})$} & \colhead{$\alpha$} & \colhead{$t_{disk}$ (Myr)} & \colhead{$t_{viscous}$ (Myr)} }
\startdata
0 &  4 & $10^{-3}$ & $\infty$ & 3.6 \\
$\frac{1}{10}$ &  4 & $10^{-3}$ & 2.2 & 3.6  \\
1 &  4 & $10^{-3}$ &  0.22 & 3.6 \\
10 &  4 & $10^{-3}$ & 0.022 & 3.6 \\
0 &  7& $10^{-3}$ & $\infty$ & 3.6 \\
$\frac{1}{10}$ &  7 & $10^{-3}$ &  2.2 & 3.6 \\
1 &  7 & $10^{-3}$ &  0.22 & 3.6 \\
10 &  7 & $10^{-3}$ &  0.022 & 3.6 \\
0 &  10 & $10^{-3}$ &  $\infty$ & 3.6 \\
$\frac{1}{10}$ &  10 & $10^{-3}$ & 2.2 & 3.6 \\
1 &  10 & $10^{-3}$ &  0.22 & 3.6 \\
10 &  10 & $10^{-3}$ &  0.022 & 3.6 \\
0 &  4 & $10^{-2}$ & $\infty$ & 0.36 \\
$\frac{1}{10}$ &  4 & $10^{-2}$ & $>5$ & 0.36 \\
1 &  4 & $10^{-2}$ & 2.2 & 0.36 \\
10 &  4 & $10^{-2}$ & 0.02 & 0.36 \\
0 &  4 & 0 & $\infty$ & $\infty$ \\
$\frac{1}{10}$ &  4 & 0 & 0.5 & $\infty$ \\
1 &  4 & 0 & 0.05 & $\infty$ \\
10 &  4 & 0 & 0.02 & $\infty$ \\
\enddata
\tablecomments{We keep $M_b$ fixed to 7 $M_{Jup}$ in every configuration. }
\end{deluxetable*}

We find that migration of the planets is significantly affected by the viscous timescale. Direct comparisons between the disk and viscous timescales are shown in Table \ref{tbl:timescales}. In the case of larger $\alpha$ ($10^{-2}$), the planets migrate outwards if $\kappa = 0$ and $\kappa = \frac{1}{10}$ (i.e., if $t_{viscous} > t_{disk}$). This does not occur in $\kappa = 1$ and $\kappa = 10$ cases, where $t_{disk} > t_{viscous}$, and we see an inward migration instead. Similarly, the nominal $\alpha$ value ($10^{-3}$) yields the same result, but with $\kappa = \frac{1}{10}$ achieving $t_{disk} > t_{viscous}$ much sooner, and therefore only displaying this outward migration in the case of $\kappa = 0$. For the case of $\kappa = \frac{1}{10}$, the planets already migrate inwards after about 0.5 Myr.
\par
In the case of $\alpha = 0$, when the disk photoevaporation is non-existent ($\kappa = 0$), both planets stay near their initial semi-major axis until the end of the simulation, migrating slightly inwards as time evolves. This migration is less efficient than in the $10^{-3}$ case. In the case where $\kappa = \frac{1}{10}$, the planets slightly migrate inwards as time evolves; however, again, that is less than in the $\alpha = 10^{-3}$ case, since the disk does not efficiently fill the large gap formed by the planets. For $\kappa = 1$ and $\kappa = 10$, we see a similar behavior, but less efficiently as $\kappa$ increases. In the $\alpha = 0$ case, we also note that the planets migrate towards each other very early in the simulations, except for the $\kappa = 10$ case, where both initially migrate inwards. The amplitude of the initial outward migration for the inner planet therefore depends directly on $\kappa$ (decreases as $\kappa$ increases). 
\par

\subsection{Long Term Stability} \label{results-stability}

\begin{figure*}
    \centering
    \centering{{\includegraphics[width= \textwidth]{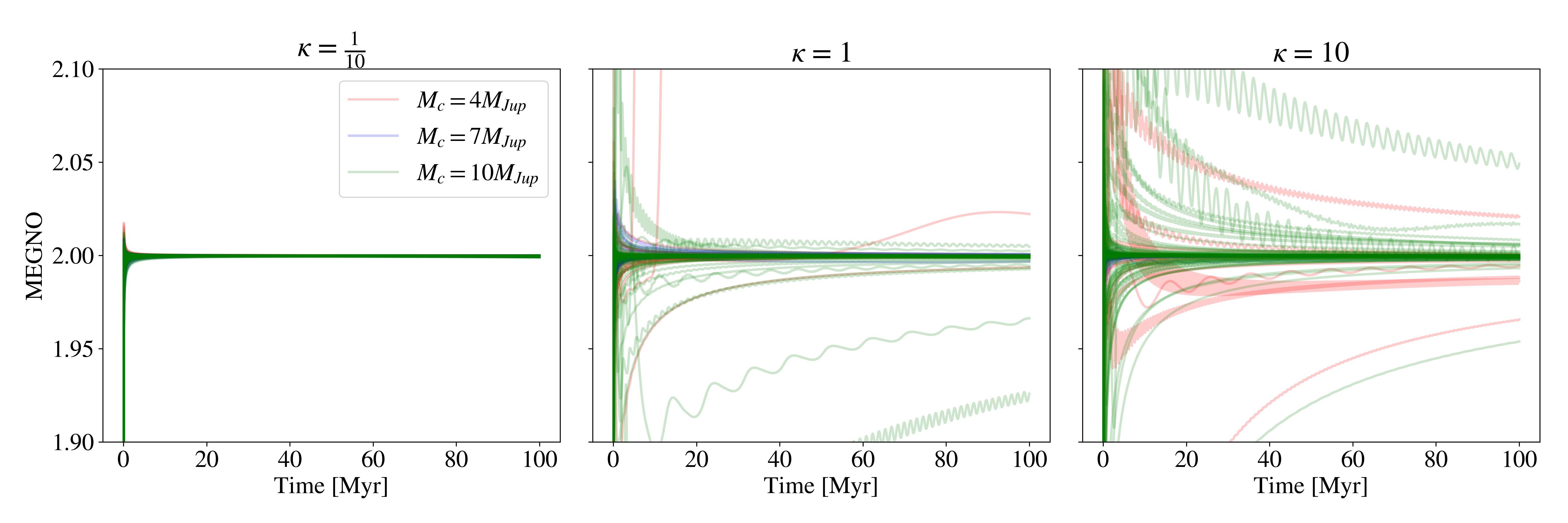} }}%
    \qquad
    \centering{{\includegraphics[width= \textwidth]{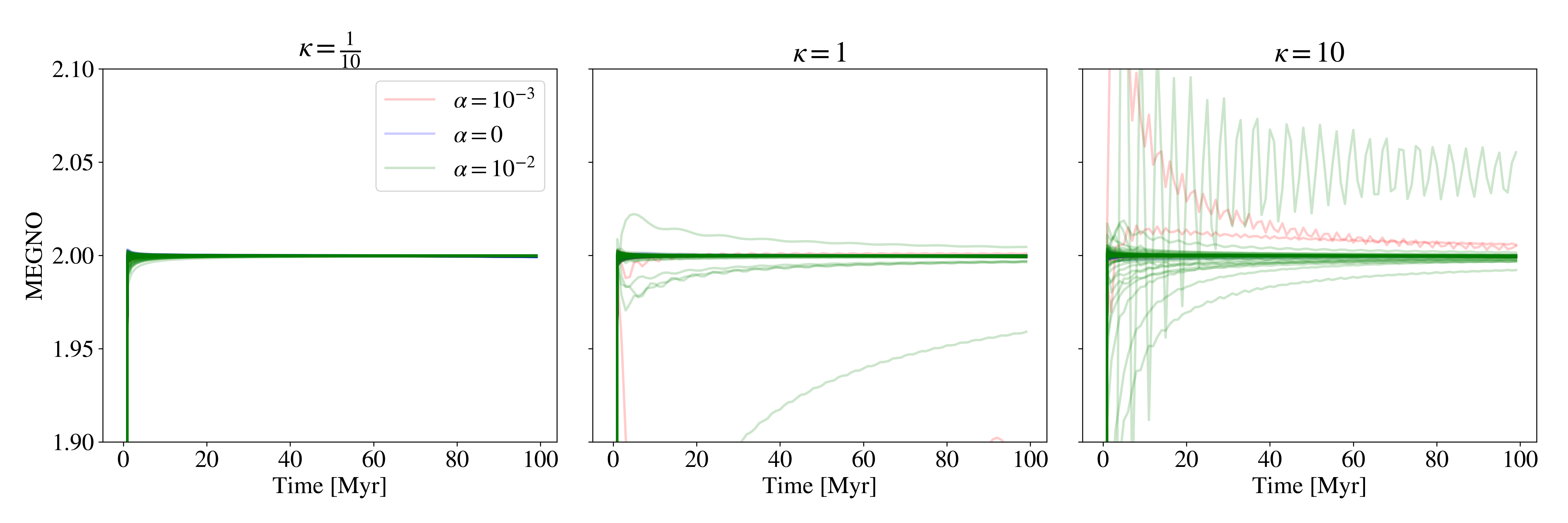} }}%
    \caption{Evolution of the system using the MEGNO chaos indicator. For stable configurations, the value must converge to $\simeq$ 2. The different panels show the MEGNO evolution for different disk photoevaporation rates $\kappa$ (different columns) and planet c masses $M_c$ (top row) and viscosity parameter $\alpha$ (bottom row).}%
    \label{fig:1gyr-nbody}
\end{figure*}

\begin{figure*}
    \centering
    \centering{{\includegraphics[width= \textwidth]{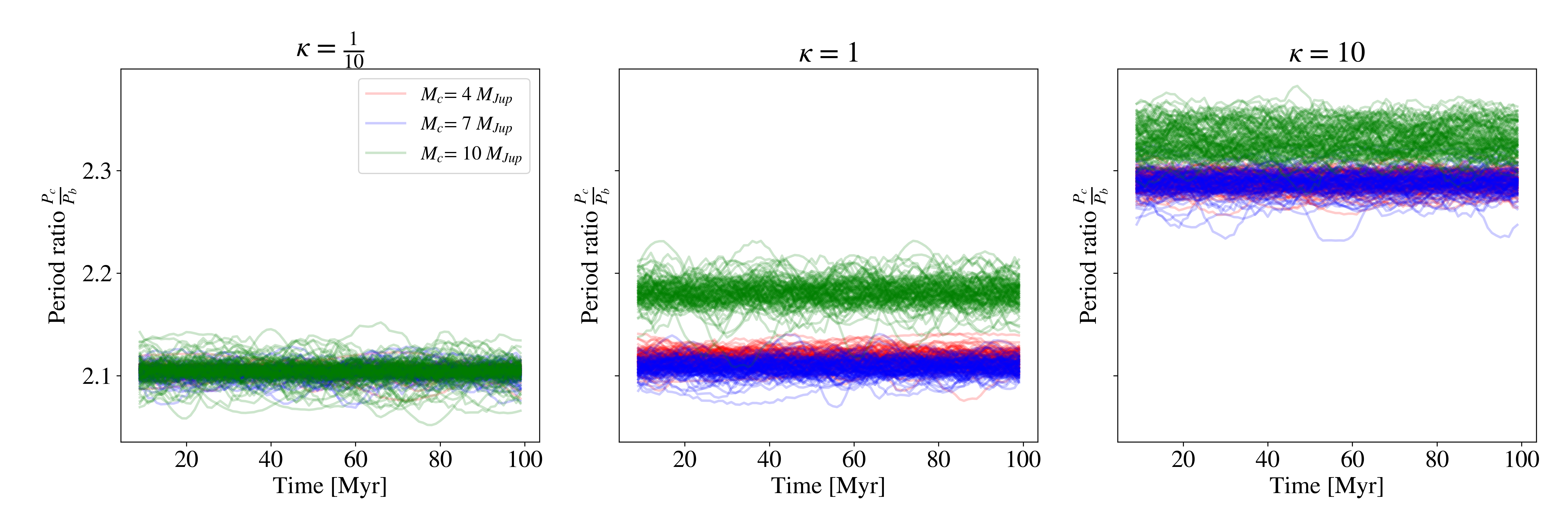} }}%
    \qquad
    \centering{{\includegraphics[width= \textwidth]{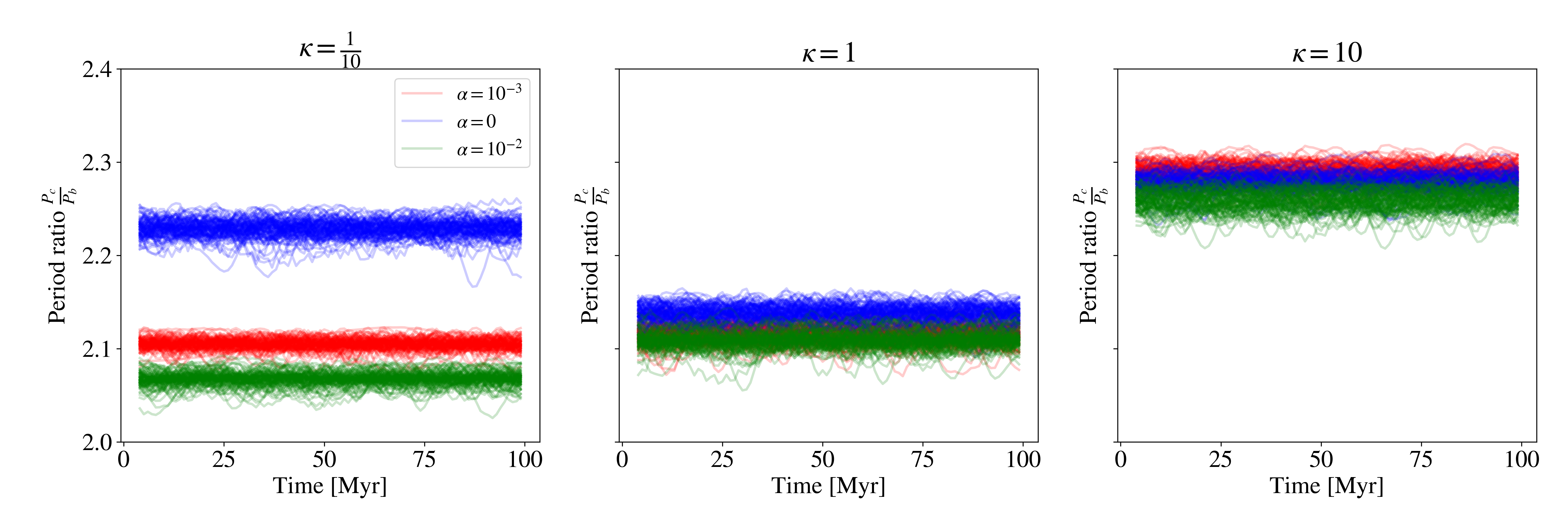} }}%
    \caption{Evolution of the system's period ratio. The different panels show the period ratio evolution for different disk photoevaporation rates $\kappa$ (different columns) and planet c masses $M_c$ (top row) and $\alpha$ (bottom row).}%
    \label{fig:1gyr-nbody-periodratio}
\end{figure*}

We perform N-body simulations for an integration time of 100 Myr, which corresponds to $\sim 10^{6}$ orbits of the inner planet and $\sim 5\times10^{5}$ orbits of the outer planet, and sufficiently long such that any instability can be captured \citep{Panichi2017}. We randomly sample 100 configurations from the final 10\% osculating elements from when the disk is fully dissipated in each $\kappa$ case. For each 100 random draws, we track the planets' orbital parameters and the MEGNO chaos indicator. Our results for the MEGNO chaos indicator are shown in Figure \ref{fig:1gyr-nbody} and summarized in Tables \ref{tbl:nbody-percent-mass} and \ref{tbl:nbody-percent-alpha}. Since the MEGNO value acts as a chaos indicator, the system remains stable for 10--100$\times$ longer than the integration time tested here (100 Myr). Therefore, for stable cases, the system remains stable for at least 1 -- 10 Gyr. We confirm this by computing the slope of the MEGNO evolution, which can be related to the Lyapunov timescale \citep{Gozdziewski2001}. The Lyapunov timescale can be approximated via the inverse of the slope (divided by 2) of the MEGNO evolution over time for chaotic systems. For regular systems, the MEGNO value converges to 2, yielding a near-zero slope and an effectively infinite Lyapunov timescale. We find that even in cases where MEGNO $>$ 2.05, there is no linear growth in MEGNO, but rather a fast growth ($<$ 5 Myr) followed by a decay. In every case, the system is not disrupted, which can be seen by the period ratio evolution plots in Figure \ref{fig:1gyr-nbody-periodratio}. An example of how leaving the MMR center can lead to the disruption of the system and consequently to shorter Lyapunov timescales is shown in Appendix \ref{lyapunov}.
\par

We find that the system is likely regular due to the resonance-locking of the planets, which can be attributed to the planetary migration in the disk. The system's stability is not significantly affected by $\alpha$. Even when the planets migrate outwards ($\alpha$ = $10^{-2}$; $\kappa$ = $\frac{1}{10}$), the system remains stable, with the resonance locking and damping of the inner planet's eccentricity assisting its long-term stability.

\begin{deluxetable}{cccc}
\tablecaption{Percent of Configurations that remain stable ($M_c$)} \label{tbl:nbody-percent-mass}
\tablewidth{20pt}
\tablecolumns{4}
\tabletypesize{\scriptsize}
\tablehead{\colhead{ $\kappa$ } & \colhead{$M_c$ = 4 $M_{Jup}$} & \colhead{$M_c$ = 7 $M_{Jup}$} & \colhead{$M_c$ = 10 $M_{Jup}$}}
\startdata
$\frac{1}{10}$  &  100\% &  100\% & 100\% \\
1  &  93\% &  100\% & 88\%   \\
10   &  89\% &  99\% & 64\%\\
\enddata
\tablecomments{We consider stability any configuration where MEGNO does not go outside of the range between 1.95 -- 2.05 in our N-body simulations. Here, $\alpha$ is $10^{-3}$.}
\end{deluxetable}

\begin{deluxetable}{cccc}
\tablecaption{Percent of Configurations that remain stable ($\alpha$)} \label{tbl:nbody-percent-alpha}
\tablewidth{20pt}
\tablecolumns{4}
\tabletypesize{\scriptsize}
\tablehead{\colhead{ $\kappa$ } & \colhead{$\alpha$ = 0} & \colhead{$\alpha = 10^{-3}$} & \colhead{$\alpha = 10^{-2}$}}
\startdata
$\frac{1}{10}$  &  100\% &  100\% & 100\% \\
1  &  99\% &  93\% & 99\%   \\
10   &  99\% &  89\% & 95\%\\
\enddata
\tablecomments{We consider stability any configuration where MEGNO does not go outside of the range between 1.95 -- 2.05 in our simulations. Here, $M_c$ is 4 $M_{Jup}$.}
\end{deluxetable}

\subsubsection{Resonance Structure}

\begin{figure*}
    \centering
    \centering{{\includegraphics[width= 0.8\textwidth]{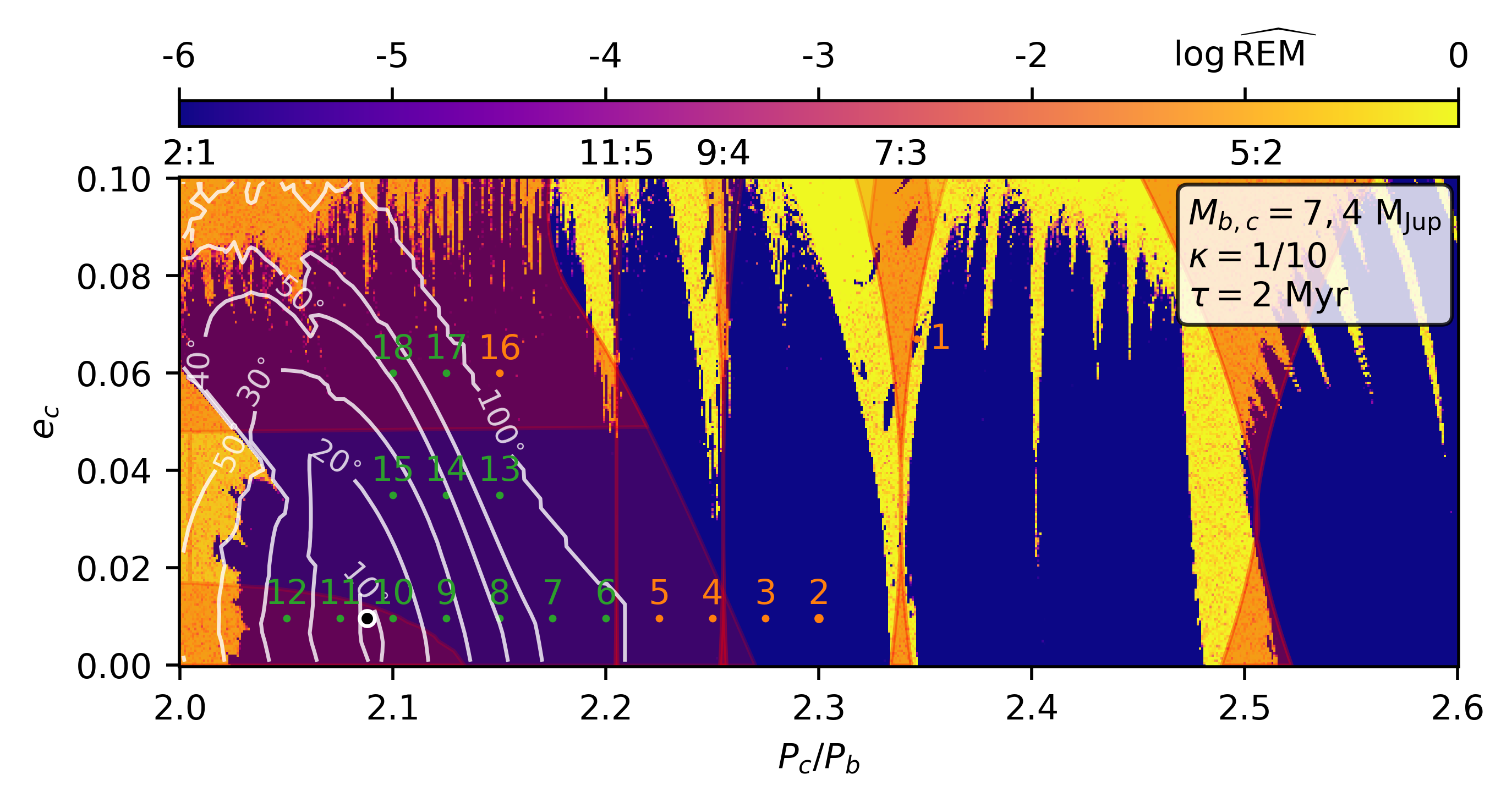} }}%
    \qquad
    \centering{{\includegraphics[width= 0.8\textwidth]{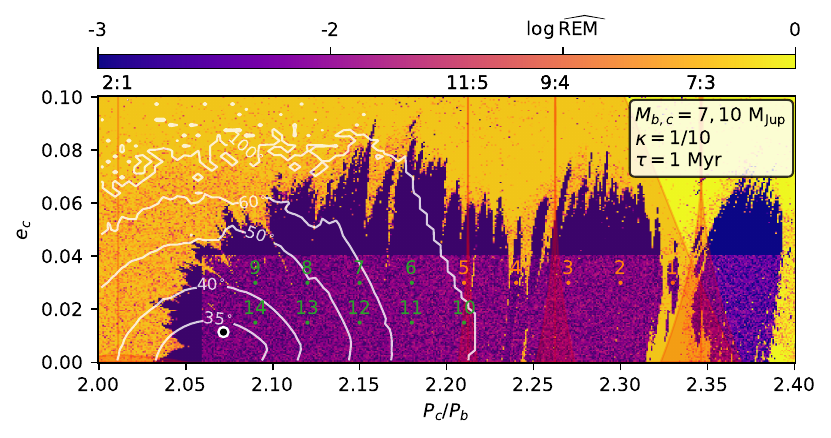} }}%
    \caption{Resonant structure of the system for the case where $M_c = 4 M_{Jup}$ (top panel) and $M_c = 10 M_{Jup}$ (bottom panel). Shaded red regions correspond to semi-analytical resonance widths obtained with the algorithm proposed by \citealt{Gallardo2021}. Purple zones correspond to numerically stable regions, where the system is expected to remain dynamically stable for 1 Myr in our numerical simulations within this parameter space. The FARGO output is marked in a black and white circle. Numbered points correspond to initial conditions for which libration of the critical angle were tested for the 2:1 resonance. Green points correspond to initial conditions with libration of one of the critical angles, while orange points correspond to libration mixed with circulation of all critical angles. The REM integrations conducted for 10,000 outermost periods, and in the rectangular, shaded region in the lower plot, for 500,000 outer orbits. White contours represent semi-amplitude of the critical angle $\theta$ for the inner planet. Note that contours for large semi-amplitudes reaching $180^{\circ}$ tend to be steep beyond $\simeq 100^{\circ}$.}
    \label{fig:res-structure}
\end{figure*}

Since the PDS 70 planets' masses are quite large, the mutual interactions between the planets likely yield a complex resonance structure. In order to visualize and confirm whether the planets are indeed inside the 2:1 mean-motion resonance (2:1 MMR), we perform an experiment to explore the neighboring regions in parameter space. We do so both numerically and semi-analytically, and \replaced{illustrate}{plot} both cases in Figure \ref{fig:res-structure} for two representative choices of the planet masses: $7, 4 M_{\rm Jup}$ and $7, 10 M_{\rm Jup}$, respectively. 

In order to assess this structure numerically, we run $N$-body simulations on a grid of selected orbital elements using the SABA4 symplectic integrator \citep{LaskarRobutel2001} and the Reversibility Error Method (REM) as a fast chaos indicator \citep{Panichi2017}. We choose this integrator and indicator because they are computationally efficient, allow for a controlled energy error, and have been shown to obtain fully equivalent results to MEGNO \citep{Panichi2017}. Using this combination allows us to obtain high-resolution dynamical maps in a computationally efficient way.
We explore the parameter space of period ratios $(\frac{P_c}{P_b}$) and the outer planet's eccentricity ($e_c$). Non-chaotic systems are expected to have low REM values, while chaotic systems have a large REM value, by a few orders of magnitude, and can be easily differentiated from stable models.
\par
Determining the resonant structure of a system with two massive planets analytically is complex \citep[e.g.,][]{Michtchenko2008a,Michtchenko2008b}. First, we used Jacobi coordinates, as astrocentric coordinates can introduce artificial variations in the osculating elements for the outer planet due to the large mass of the inner planet \citep{Gallardo2021, Zurlo2022}. Because a system with two planets and a star has six degrees of freedom, \citealt{Gallardo2021} derives a semi-analytical model that simplifies this problem to two degrees of freedom for two planets near a resonance, by fixing the longitude of ascending node (which in our case, is already zero, since we consider a coplanar system) and argument of periastron, and then computing the average of the resonant terms of the disturbing function. For the PDS 70 planets, we fix the inner planet's location and search for outer planet locations that yield a MMR. 
To determine the position and width of the most prominent and expected MMRs in the region of interest, we used the publicly available \texttt{Plares} code by \citep{Gallardo2021}, in their \texttt{version1b} case. It should be noted that the mass range in our model implies large widths of the 2:1~MMR, that can affect validity of the simplified averaging approach. Nevertheless, it gives useful insight into the qualitative features of the system. \cite{Gallardo2021} introduce a spacing parameter $\sigma$ through the tolerance for crossing of the orbits $\sigma R_{\rm H}$, where $R_{\rm H}$ is their mutual Hill radius. They recommend $\sigma=3$ for stable widths of the MMRs, and $\sigma=0.1$ for reasonably full widths. As we can see below, it becomes important especially for the 2:1~MMR.

\par
In order to compare the numerical and semi-analytical results, we plot them in dynamical maps (Figure \ref{fig:res-structure}) for $\kappa=1/10$ and $M_c = 4 M_{\rm Jup}$ and $M_c = 10 M_{\rm Jup}$, with the semi-analytical case being shown as a red shaded area for different identified resonances. We also test the libration of the two critical angles of the 2:1 MMR for different initial conditions (ICs) numbered at locations in the maps and marked with colored labels. In both the maps, the nominal initial conditions are selected with temporal $\Delta\varpi \simeq 0^{\circ}$ for epochs $\simeq 1$\,Myr and are marked with larger filled circles. The identified MMRs between the planets are marked at positions identified through the semi-analytic averaging. It turns out that the stable region detected with the REM is confined to small eccentricity $e_{\rm c}<0.1$. Identification of the 2:1 MMR width appears to be non-intuitive, as the resonance center for $P_{\rm c}/P_{\rm b} = 2$ appears unstable and we find that stable region with $P_{\rm c}/P_{\rm b} > 2$, where the MMR can be expected, has no clear border (separatrix). Moreover, for smaller outer mass (top panel in Figure  \ref{fig:res-structure}), the analytic model indicates two disjoint sets (darker shade) for $\sigma=3$, and an extended, continuous area overlapping with 11:5~MMR and 9:4 MMR. Furthermore, for the larger outer mass, the analytic prediction for $\sigma=0.1$ overlaps with even the 7:3~MMR.
\par
We can shed some light on this issue through inspecting critical angles for the selected ICs. For the smaller mass, we detected clear libration of $\theta_{2:1,2}$ (green labels) around $180^{\circ}$ with amplitude becoming smaller as we approach $P_{\rm c}/P_{\rm b} \simeq 2$. In some cases, we can see both angles librating. In the transition zone, near the analytic MMR border, the libration interchanges with circulation. For ICs=2,3 there is essentially no libration. Moreover, IC=1 can be clearly attributed a libration of one of the  critical angles of the 7:3~MMR, confirming the semi-analytical averaged model and the REM indicator completely agree. As mentioned, for larger outer mass, the MMR structure is even more extended and diffuse in the sense of the critical angles but the general view is the same. To interpret the results of these experiments, we can invoke the MMR resonance overlap which is strong for the large masses. Given the higher order MMRs are identified, we can explain the global instability region just above $e_{\rm c} \simeq 0.1$ (especially for the larger mass case) emerging due to the overlap of these resonances. Due to the mutual perturbation and the MMRs overlap, the 2:1 MMR has a wide diffuse border. The regions, which appear as stable for relatively short REM integration time 10,000 $P_c$ \added{do not change their status after much longer time, $5 \times 10^5 P_c$, either. This is illustrated in Figure \ref{fig:res-structure} (lower panel) as a rectangular, shaded region. Interchanged libration and rotation of the critical angle in the intermediate zone should not be confused with a signature of separatrix chaos. The solutions marked with dark colours remain stable, and the alternating evolution of one of the critical angles is due to the vicinity of the wide 2:1~MMR.}. 
\par
We conclude that the MMR structures in the ($P_{\rm c}/P_{\rm b},e_{\rm b}$) plane are complex. A more effective approach to detecting separatrices within the MMR would be the use of representative planes of initial conditions, constrained by integrals of motion \citep[e.g.,][]{Michtchenko2008a}. Unfortunately, our attempt to apply this method also did not yield a clear dynamical image of the resonance.

\par

\subsection{Orbit Fits and Stability}

We perform updated orbit fits for PDS 70 b and c given the publication of new relative astrometry data for both planets. Our goal is to compare the stability rate of the hydrodynamical simulations (i.e., theoretical dynamical evolution that considers the effects of the disk) with that obtained using the current orbit fits coupled with N-body simulations for both planets. The latter is a common means of dynamical assessment in the literature for multi-planet systems \citep{Gozdziewski2009, Wang2018, Wang2021, Thompson2023, Hinkley2023, Sappey2025}. \par
The astrometry used includes points from 2012 -- 2018 (\citealt{Christiaens2019}, \citealt{Keppler2018}, \citealt{Haffert2019}, \citealt{Muller2018} and \citealt{Wagner2018}), as well as more recent 2021 -- 2022 data from \citealt{Wahhaj2024} and 2022 -- 2024 from \citealt{Close2025}. We also include a new relative radial velocity point (RV) for PDS 70 b from \citealt{Hsu2024}, which can provide third-dimensional information for the orbit fit \citep{DoO2023, DoO2024}.
For our orbit fits, we use the \texttt{octofitter} orbit fitting package \citep{Thompson2023}. We use observable-based priors in all of our fits \citep{ONeil2019}, which aim to decrease biases in orbit fits where the data spans a short orbital arcs. We set the system parallax ($\pi$) to be the Gaia DR3 measurement of $8.8975 \pm 0.0191$ mas and the system mass to be 0.85 $\pm$ 0.10 \Msol. \par
We test two orbit fits with the new data: one where the planets' orbital planes are unconstrained and one where they must share the same orbital plane (coplanar). We perform joint orbit fits, which include the epicycle approximation to account for the inner planet's gravitational effect on the outer planet \citep{Lacour2021}. For our coplanar orbit fit, we set the inclination prior to be a sine prior for the entire system, and set the $\Omega$ prior to be uniform between 0 and 360$^{\circ}$. The inclination results for both cases are consistent with each other; i.e. the planets are likely in a near-coplanar configuration. Our results are presented in Table \ref{tbl:orbits}.\par
We find that the orbit fits for b and c are consistent with the possible outcomes from the hydrodynamic simulations. Specifically, the ranges in semi-major axis and eccentricity encompass the solutions presented in Sections \ref{planetparam-mckappa} and \ref{alpha-results}.

\begin{deluxetable}{ccc}
\tablecaption{Orbit Fit for PDS 70 b and c} \label{tbl:orbits}
\tablewidth{20pt}
\tablecolumns{3}
\tabletypesize{\scriptsize}
\tablehead{\colhead{Parameter} & \colhead{Unconstrained} & \colhead{Coplanar}}
\startdata
$a_b$ (AU)  & $22\substack{+5 \\ -4}$ & $21.2\pm 2.8$ \\
$e_b$  & $0.26\substack{+0.25 \\ -0.20}$ & $0.25\substack{+0.19 \\ -0.18}$ \\
$i_b$ (°)  & $131\substack{+18 \\ -10}$ & $130\substack{+11 \\ -7}$ \\
$\Omega_b$ (°)  & $258\substack{+94\\ -247}$ & $338\substack{+12 \\ -47}$ \\
$\omega_b$ (°)  & $206\substack{+127 \\ -175}$ & $294\substack{+43 \\ -208}$ \\
\hline
$\theta_b$ (°)  & $133.3\pm1.5$  & $133.6\pm1.5$ \\
$a_c$ (AU) & $29\substack{+10 \\ -6}$ & $33\substack{+9 \\ -7}$ \\
$e_c$  & $0.15\substack{+0.25 \\ -0.12}$ & $0.18\substack{+0.21 \\ -0.14}$ \\
$i_c$ (°)  & $132\substack{+14 \\ -9}$ & $130\substack{+11 \\ -7}$ \\
$\Omega_c$ (°)  & $180\substack{+153 \\ -144}$ & $338\substack{+12 \\ -47}$ \\
$\omega_c$ (°)  & $175\substack{+107 \\ -138}$ & $206\substack{+55 \\ -53}$ \\
$\theta_c$ (°)  & $270.6\pm 0.8$ & $270.4\pm0.7$ \\
\hline
$M_{\mathrm{sys}}$ ($M_\odot$)  & $0.93\pm 0.09$ & $0.93\pm 0.09$\\
$\pi$ (mas) & $8.898\pm 0.020$ & $8.89\pm 0.02$\\
\enddata
\tablecomments{$\theta_b$ and $\theta_c$ correspond to the position angle at the reference epoch (MJD 60000). The values listed here represent the median and 68\% credible interval for each parameter.}
\end{deluxetable}

We then evaluate the long-term stability of the system (over 100 Myr) using the orbital posteriors and the MEGNO indicator. We draw from the orbital posterior families and test the cases where planet b has a mass of 7 $M_{Jup}$ and planet c has a mass of 4, 7 and 10 $M_{Jup}$. We find that only 4.2 -- 4.5\%  of orbital posterior configurations are stable in the coplanar case (depending on $M_c$'s mass), with most configurations resulting in one of the planets being ejected ($>$ 500 AU). For the non-coplanar case, that $<$ 1\% of the configurations remain stable. The mass of c does not strongly affect stability (this was similarly found by \citealt{Wang2021}), however, it does affect which planet is ejected (the lighter one). 
\par
\begin{deluxetable}{ccc}
\tablecaption{Percent of Configurations that remain stable for at least 100 Myr (2 Planet Orbit Fit)} \label{tbl:nbody-percent-mass}
\tablewidth{20pt}
\tablecolumns{3}
\tabletypesize{\scriptsize}
\tablehead{\colhead{ $M_c$ ($M_{Jup}$) } & \colhead{Unconstrained} & \colhead{Coplanar}}
\startdata
4   &  0.9\% &  4.2\% \\
7   &  0.9\% &  4.5\%  \\
10   &  0.7\% &  4.2\% \\
\enddata
\tablecomments{We consider stability any configuration where MEGNO does not go outside of the range between 1.95 -- 2.05 in our simulations.}
\end{deluxetable}
\section{Discussion} \label{discussion}

\subsection{Planet Migration and Resonance Locking}
\begin{figure}
  \begin{center}
\centerline{\includegraphics[width=0.45\textwidth]{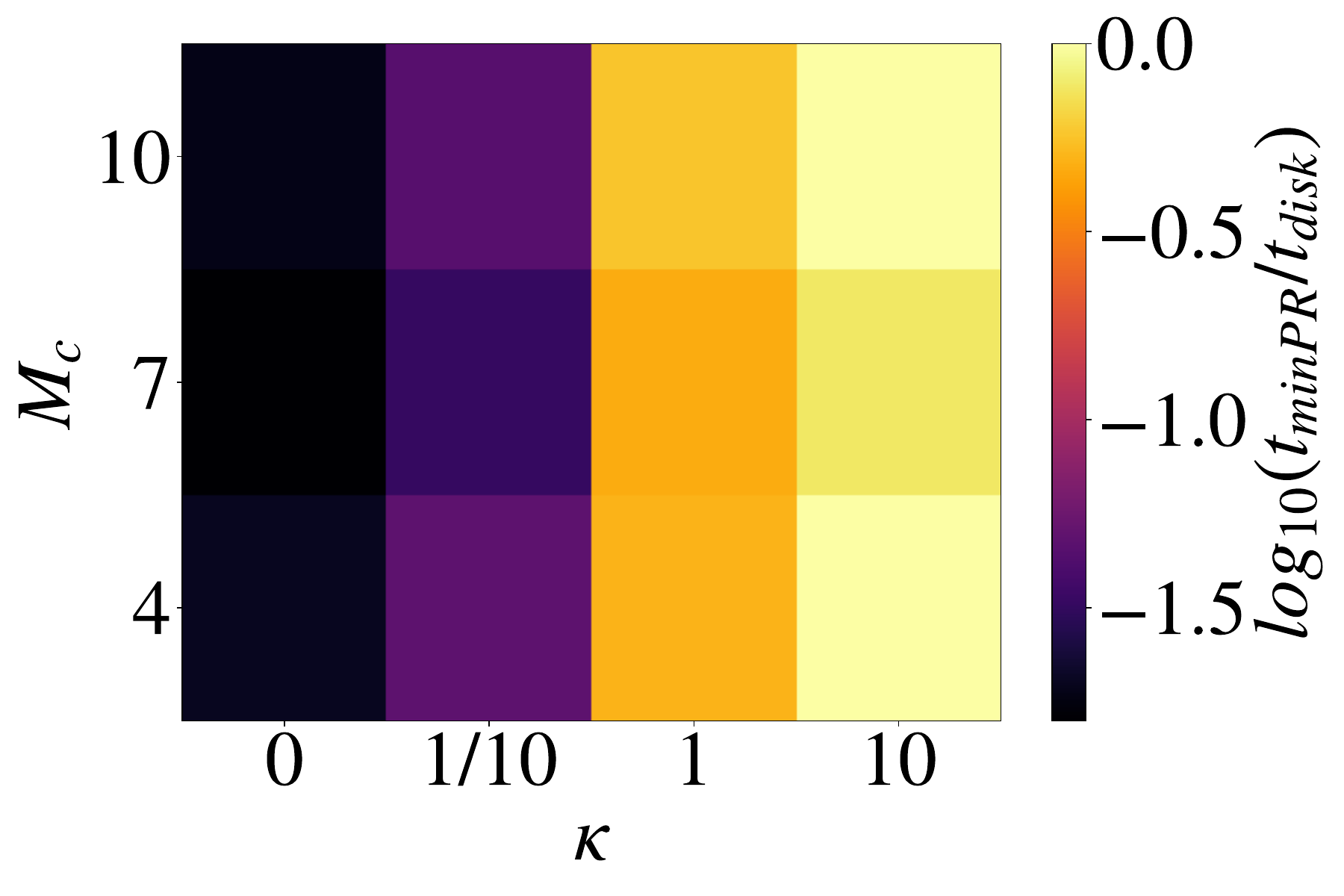}}
\caption{Minimum period ratio time scales for each case divided by the disk life times. We find that the the system is less likely to meet the center of the resonance location before the disk dissipates if these timescales are comparable, which shows why the resonance locking is stronger if the disk is present for a longer time.}
\label{fig:minpr-disklifetime}
  \end{center}
\end{figure}

The planets migrate in the disk due to the inner/outer spiral arm torques and the co-rotation torques \citep{GoldreichTremaine1980}. In order to get locked into resonance, the planets must go through convergent migration (i.e. their period ratios must decrease over time) and meet at a resonance location (e.g. period ratios of 2:1, 5:2, etc). We see this convergent migration in all cases for the PDS 70 system in our simulations (period ratios decrease over time), which occurs due to the formation of a common gap in the disk. However, the convergent migration must last long enough such that the planets actually arrive at the center of the resonance location (i.e. their period ratio must meet $\sim$ 2:1 before the disk dissipates around the planets). Not all simulations meet this criterion; in particular, high $\kappa$ values do not yield this result. In order to visualize this, we plot the minimum period ratio timescale divided by the disk lifetime timescale for each case, and find that, indeed, $\kappa = 10$ values are less likely to yield planets that arrive at a minimum period ratio that is near the 2:1 resonance, because the disk lifetime is comparable ($\geq$ 1) to the minimum period ratio timescale. This is visualized in Figure \ref{fig:minpr-disklifetime}. \par
This timescale comparison only tells us whether the planets will meet the resonance location of 2:1. In order for them to \textit{stay} there, they must meet the adiabatic criterion \citep{Malhotra1993, Masset2001, Goldreich2014, Batygin2015}, which requires that they migrate slowly enough to get trapped into the resonance. In other words, the resonant timescale (mean motion timescale) must be shorter than the migration timescale. The adiabatic criterion for a 2:1 mean-motion resonance where the outer planet is denoted as c and the inner planet is denoted as b reads \citep{Malhotra1993}:
\begin{equation}
    \frac{|\dot{a_c|}}{a_c\Omega_c'} << 3q_be_c,
\end{equation}
where $a_c$ is the semi-major axis of planet c, $q_b$ denotes the mass ratio between planet b and the star, and $e_c$ denotes the eccentricity of planet c. We compute this adiabatic criterion where the planets approach the 2:1 commensurability in each case, and find that in all cases the system meets the adiabatic criterion by at least two orders of magnitude. Therefore, the planets get locked into the 2:1 resonance upon approaching its location. The presence of the disk facilitates (and in fact, ensures) this approach. In other words, likely due to the low disk density and the proximity to the 2:1 MMR in the beginning of the simulations, the planets migrate slowly enough that they get locked into the 2:1 resonant state rather than simply passing through it. The only case in which the planets do not reach the center of the 2:1 MMR is if the X-ray luminosity of the star is large enough to remove the disk material that would lead them to migrate in the first place, as is seen in all cases where $\kappa = 10$. Therefore, it appears that stars with large X-ray luminosities can dissipate the disk before full convergent resonant migration occurs. Whether the planets' migration \textit{after} resonance locking is inwards or outwards is dependent on the planets' mass ratio and disk viscosity, in particular because their migration becomes coupled once resonance locking occurs \citep{Masset2001}.

We note that the period ratio of the planets is not exactly an integer -- they generally ``lock" into ratios that are slightly larger than 2:1, at about 2.1:1 (see Figure \ref{fig:res-structure}). The disk migration is the origin of this effect. In short, the gas-disk interaction leads to the precession of the planets' longitude of pericenter, which in turns slightly shifts the location of the resonance \citep{Tamayo2015} to slightly larger values. This effect likely assists in the stabilization of the system by preventing the planets from going through resonant overlaps, which can be destabilizing \citep{Morrison2016}. Transiting planets found with Kepler also have this slight shift from perfectly integer ratios \citep{Goldreich2014}.


\subsubsection{Outer planet mass dependence} 

The planets carve a common gap in the disk, which can be seen in Figure \ref{fig:densplots}. This leads the inner Lindblad torque of the outer planet and the outer Lindblad of the inner planet to become suppressed. The remaining components that could lead to significant migration are the inner Lindblad torque of the inner planet (planet b) and the outer Lindblad torque of the outer planet (planet c). If the disk does not evaporate ($\kappa = 0$) and $M_c < M_b$, the planets migrate outward after resonance locking because the inner Lindblad of the inner planet is larger. This is known as the Masset and Snellgrove mechanism \citep{Masset2001}, and has been reported in other simulations involving gas giants \citep{Crida2009}, including the Grand Tack Model of our own Solar System simulating Jupiter and Saturn's early migration \citep{Walsh2011}. The opposite occurs if $M_c > M_b$, as the outer Lindblad torque of planet c will promote a larger torque than the inner Lindblad torque of planet b. This is clearly illustrated in Figure \ref{fig:sma-mass}. The reason that the outer migration is stronger in the case where $M_c = M_b$ is that the common gap is formed sooner due to the larger outer planet mass (see Figure \ref{fig:densplots}, top panels left and middle figures), so this outward migration takes effect sooner. \par
If the disk evaporates, however, the migration is inwards in all cases. Given the disk's photoevaporation profile (Figure \ref{fig:photoevaporationpng}), the inner disk is more efficiently (4--5x) depleted than the outer disk, so the inner Lindblad of the inner planet gets depleted sooner than the outer Lindblad of the outer planet. Therefore, the outer Lindblad torque dominates, and the planets migrate inward. The strength of this migration is inversely proportional to the photoevaporation rate; for a larger $\kappa$ the migration rate is smaller.\par
The eccentricity excitation of both planets occurs upon resonance locking. The inner planet's excitation is also dependent on the mass of the outer planet. We conclude this is due to the periodic perturbations of both planets at conjunctions, as the inner planet meets the outer planet periodically. \citealt{Goldreich2014}'s equation 3 shows this relationship, derived from Lagrange's equation of motion to first order in eccentricity for the inner planet:
\begin{equation}
    \dot{e} = \beta q'nsin\phi - \frac{e}{\tau_e}
\end{equation}
i.e., the eccentricity evolution of the inner planet is coupled to the mass ratio of the outer planet to the star ($q'$). Here, $\tau_e$ denotes the eccentricity damping timescale,$\phi$ is the dominant term in the inner planet’s disturbing function has resonant argument, and n is the order of the resonance. \par

\subsubsection{Viscosity dependence} 
When the disk does not dissipate at all ($\kappa=0$), the planets migrate outward. This is because in our case the inner planet is more massive than the outer planet for the initial conditions that vary the disk's viscosity. The outward migration relative to planet mass is discussed in the previous section. The efficiency of this outward migration clearly depends on $\alpha$ -- if $\alpha$ is larger, the disk is more viscous; and the gap carved by the planet is more efficiently refilled. This leads the co-rotation torque to become significant.  \par
The depth and width of the gap carved by a giant planet relies on the balance between the Lindblad torques, which open the gap, and viscous torques, which close the gap \citep{GoldreichTremaine1980, FungShiChiang2014}. In principle, any planetary mass can open a gap in the disk, with larger masses leading to deeper and wider gaps, especially in less viscous disks (e.g. \citealt{FungShiChiang2014, Duffell2015}).
\par
\begin{figure*}
  \begin{center}
\centerline{\includegraphics[width=\textwidth]{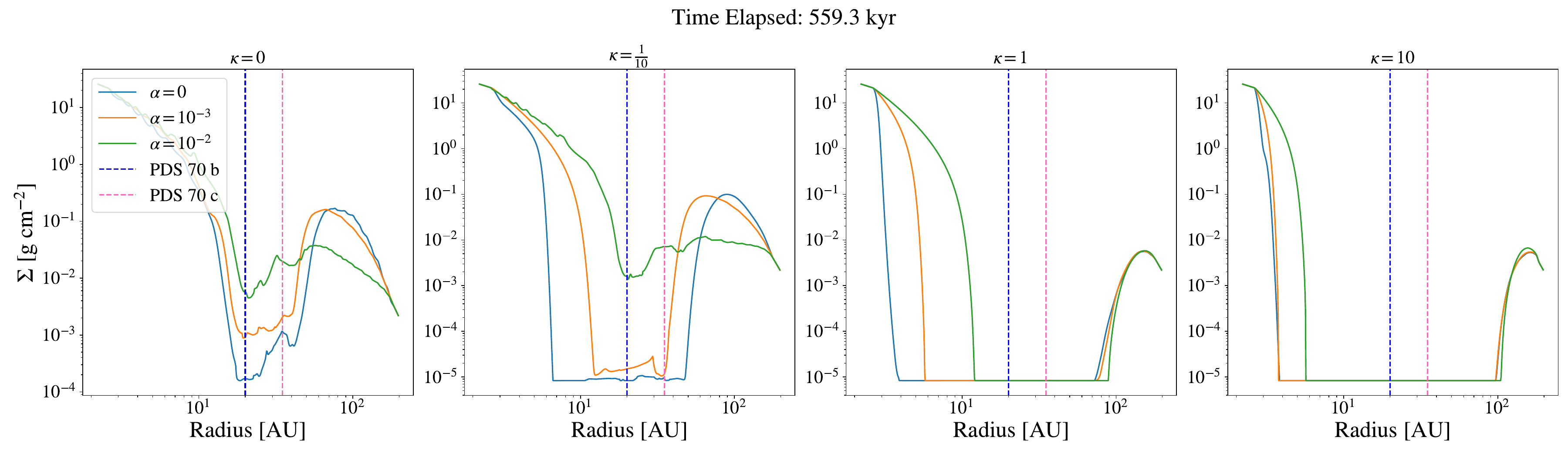}}
\caption{Disk density for different $\alpha$ and $\kappa$ values at t $\sim$ 0.6 Myr. Note how the gap is not as deep for larger $\alpha$, while it is significantly depleted for smaller $\alpha$. In the $\kappa$ = 1/10 case, the gap depth varies by about 3 orders of magnitude.}
\label{fig:dens-alpha}
  \end{center}

\end{figure*}

When the disk dissipates slowly ($\kappa = \frac{1}{10}$), the planets migrate outwards only in the alpha $10^{-2}$ case. The viscous timescale is on the order of a few Myr for the $10^{-3}$ case, a few hundreds of thousands of years for the $10^{-2}$ case, and very large for $\alpha$ = 0 (i.e., the gap gets depleted rapidly, and the disk is not actively trying to refill it). We showcase this in Figure \ref{fig:dens-alpha} for a snapshot at t $\sim$ 0.6 Myr. The gap is more efficiently carved for smaller $\alpha$, which leads to a rapid saturation of the co-rotation torque and the weakening of the Lindblad torques on the planets.

The reason the planets do not migrate outwards for $\alpha$ of $10^{-3}$ and 0 is because the viscous timescale is larger than the lifetime of the disk. Therefore, the outer Lindblad resonance of the outer planet dominates the migration pattern, causing the planets to migrate inward. This inward migration effect has been recently reported in other hydrodynamic simulations that explore low-viscosity disks \citep{Griveaud2023}.
\par
For the case of $\kappa = 1$, the disk is dissipated within a few hundreds of thousands of years. The planets are locked into resonance and their eccentricities are fairly low ($<$ 0.1 for both). Both planets migrate inward, with the inner planet migrating slightly more than the outer one. The initial migration causes the inner planet to move outward slightly from its initial location of 20 AU, and inward slightly for the outer planet from its location at 35 AU. This is due to the planets locking into resonance, but this migration is less efficient for smaller $\alpha$. Once the planets are locked into resonance, they both migrate inwards. This is due to the lack of inner disk material: the photoevaporation is more efficient at removing material from the inner region of the disk, causing the migration to be inward. That is more efficient for larger $\alpha$. In the case of $\kappa = 10$, the planets do not get locked into resonance, as the disk dissipates within a few tens of thousands of years. They both migrate inward. That migration is stronger for larger $\alpha$ and for the inner planet. Being more massive, the inner planet creates a gap first, depleting the inner Lindblad torque and corotation torques that would cause it to migrate outward, but not the outer Lindblad region that would cause it to migrate inward.
\par

\subsection{PDS 70 d?} \label{results-pds70d}

Beyond the confirmed protoplanets PDS 70 b and c, recent works have identified possible point sources that could be inner protoplanets embedded in the disk. Specifically, \citet{Mesa2019} pointed out a third object located at about 0.12'' from the star with VLT/SPHERE, which would place its semi-major axis at about 13.5 AU. Coupling the contrast of the object of 7.27$\times$ $10^{-5}$ with the assumption that the dust is filling the Hill radius around the planet, they find that the lower limit on this candidate's mass is $\sim$ 17.3 $M_{\oplus}$. This location is near the 1:2:4 mean-motion resonance (MMR) location for a third inner planet, which would be at about $\sim$ 12.6 AU. Most recently, \citet{Christiaens2024} used JWST/NIRCam to image the PDS 70 system. They find that the candidate appears to move in a Keplerian motion at 13.5 AU. Using nine years of imaging data from VLT/SINFONI, VLT/NaCo, VLT/SPHERE, and JWST/NIRCam, \citealt{Hammond2025} found that the third potential protoplanet exhibits Keplerian motion, though its nature as a protoplanet or disk feature (e.g. heated circumplanetary dust, variability) remains uncertain. This candidate, if real, would make this system's architecture similar to HR 8799, with at least three massive planets near a Laplace resonance \citep{Gozdziewski2014,Thompson2023}. \par
\begin{figure*}
    \centering
    \centering{{\includegraphics[width= 0.4\textwidth]{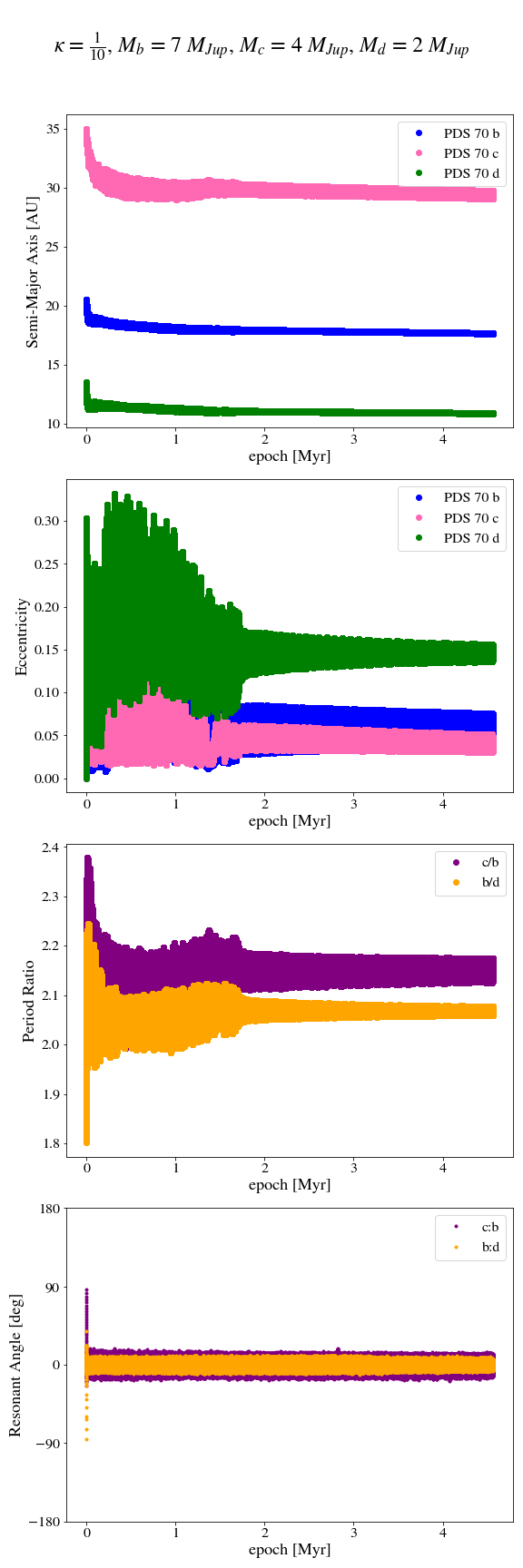} }}%
    \qquad
    \centering{{\includegraphics[width= 0.4\textwidth]{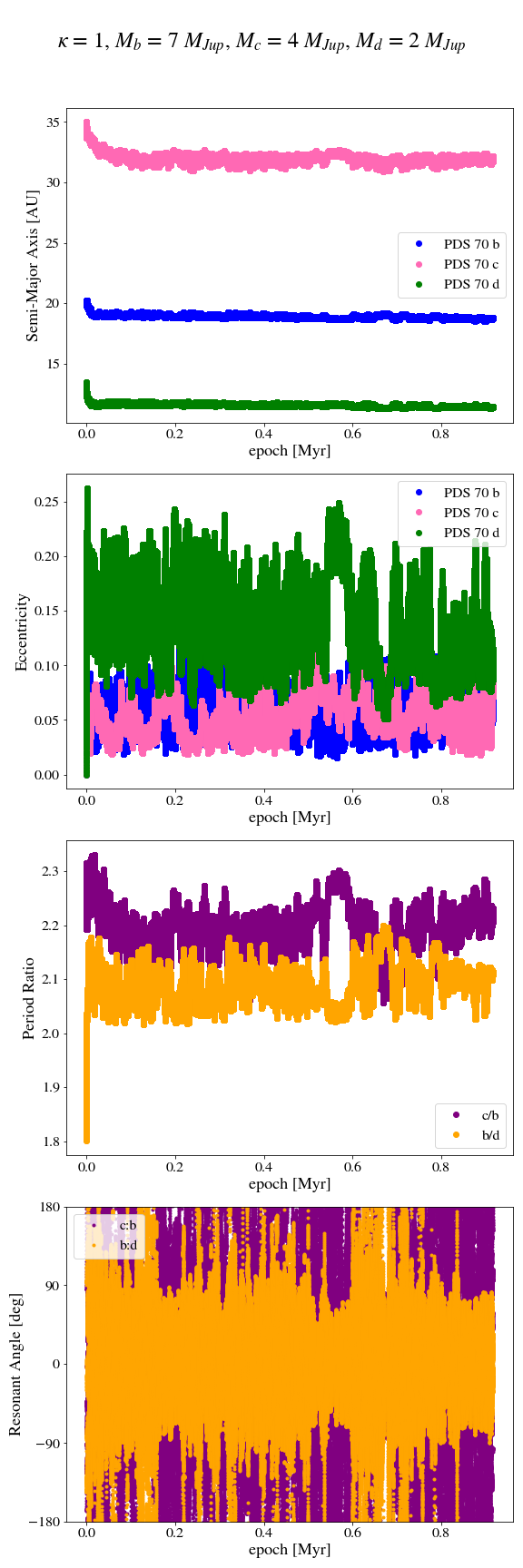} }}%
\caption{Planet parameter evolution for PDS 70 b, c and ``d" in a photoevaporating disk with $\kappa = \frac{1}{10}$ (left) and $\kappa = 1$ (right). The planets get locked into a Laplace resonance within 0.1 Myr for $\kappa = \frac{1}{10}$, but fail to migrate fully into the resonance for $\kappa$ = 1.}
   \label{fig:pds70d-fargo}
\end{figure*}

Given the likely long-term stability of the PDS 70 system and the wide separations of the two forming gas giants, it is possible that inner planets exist and are dynamically stable. We test whether a candidate, ``PDS 70 d" could be present and allow for system stability. We only test two cases for this with Dusty FARGO, where $\kappa$ is $\frac{1}{10}$ and 1, such that the disk takes longer than 0.1 Myr to dissipate.  We set the initial semi-major axis of ``PDS 70 d" to 13.5 AU, in agreement with candidate orbit fits (see \citealt{Mesa2019, Christiaens2024}) and the mass to 2 $M_{Jup}$, which is above the lower limit of 90 $M_{\oplus}$ ($\approx$ 0.05 $M_{Jup}$) and consistent with a gas giant mass. Planets b and c have masses of 7 and 4 $M_{Jup}$. The results are presented in Figure \ref{fig:pds70d-fargo}. In the case of $\kappa$ of $\frac{1}{10}$, the planets get locked into a Laplace (1:2:4 resonance) within 0.1 Myr, and all planets migrate inwards from their initial locations (average semi-major axis of $a_b$ $\sim$ 18 AU, $a_c$ $\sim$ 30 AU, and $a_d$ $\sim$ 11 AU). The eccentricity of the inner planet gets excited to $\sim$ 0.15. The eccentricity of b is slightly higher than c, with values of $\sim$ 0.06 and 0.04 respectively.
The result is significantly different in the $\kappa = 1$ case: with the disk's rapid dissipation ($<$ 0.5 Myr), the planets do not fully migrate to the center of the mean-motion resonance, evident from the changes in libration/circulation of the resonance angle and significant perturbations on the planets' period ratios.

When we integrate the system for 100 Myr to assess the MEGNO chaos indicator, in the same way we did for the PDS 70 b and c-only case. We find that the system's stability drops to 34\% for $\kappa = \frac{1}{10}$, compared to nearly 100\% for the two-planet case. Given the instability of the 3-planet case, it becomes straightforward to compute the Lyapunov timescale using the average slope of the MEGNO evolution. We obtain $t_{Lyapunov}$ of 6 Myr for $\kappa = \frac{1}{10}$ and 0.004 Myr for $\kappa = 1$, confirming that the presence of the disk leads to a stabilization of the system by allowing for planetary migration towards the center of the MMR.
\par
\begin{figure*}
  \begin{center}
\centerline{\includegraphics[width=\textwidth]{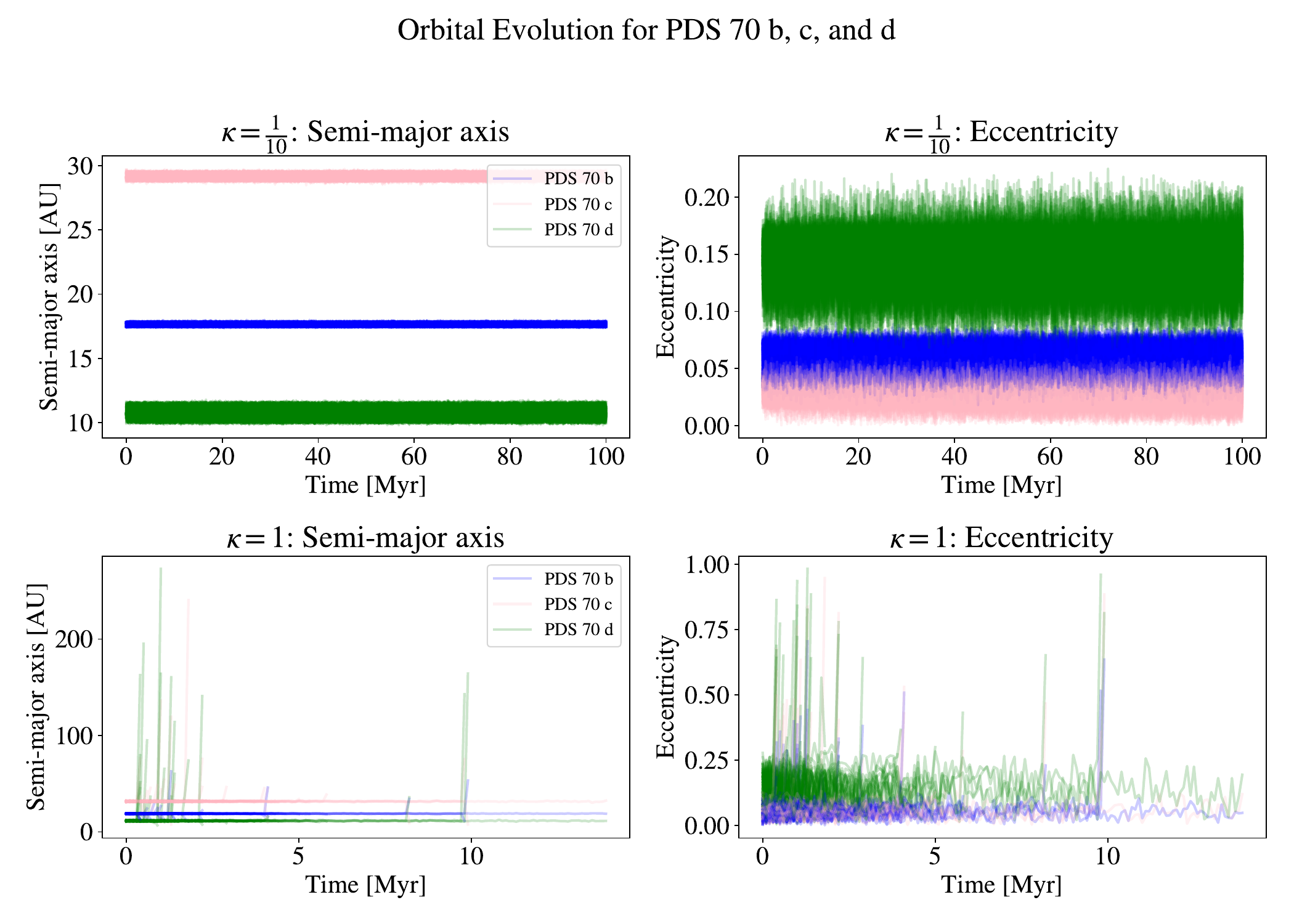}}
\caption{The PDS 70 b, c and d orbital evolution for $\kappa = \frac{1}{10}$ and $\kappa = 1$. The system becomes unstable, with ejections and scattering, in the higher photoevaporation case where the planets do not have enough time to migrate to the center of the MMR before the disk dissipates.}
\label{fig:pds70d-nbody}
  \end{center}

\end{figure*}

Although this is a significant decrease in the likelihood of stability from the 2-planet case, we note that having lower masses could potentially lead to more configurations that are likely to be stable. 
Since the uncertainties on the masses of all planets are large, it is therefore possible that the system is indeed long-term stable with three gas giant planets. The higher photoevaporation ($\kappa = 1$) has important consequences for the long-term stability of the system, with every configuration disrupting within a few tens of Myr. PDS 70 ``d", being the lightest of the three, is the most likely to get ejected or scattered to higher eccentricities and distances from the host star. This is clear in Figure \ref{fig:pds70d-nbody}. 

\begin{deluxetable}{ccc}
\tablecaption{Orbit Fit for PDS 70 b, c, and d} \label{tbl:orbitswithd}
\tablewidth{20pt}
\tablecolumns{3}
\tabletypesize{\scriptsize}
\tablehead{\colhead{Parameter} & \colhead{Unconstrained} & \colhead{Coplanar}}
\startdata
$a_b$ (AU)  & $22\substack{+5 \\ -4}$ & $20\substack{+3 \\ -2}$  \\
$e_b$  & $0.28\substack{+0.24 \\ -0.22}$ & $0.33\substack{+0.12 \\ -0.13}$ \\
$i_b$ (°)  & $132\substack{+18 \\ -11}$ & $134\substack{+6 \\ -5}$ \\
$\Omega_b$ (°)  & $287\substack{+64 \\ -272}$ & $330\substack{+16 \\ -161}$ \\
$\omega_b$ (°)  & $218\substack{+113 \\ -186}$ &  $289\substack{+41 \\ -148}$ \\
$\theta_b$ (°)  & $133.3\pm1.5$  & $133.2\pm1.4$ \\
\hline
$a_c$ (AU) & $29\substack{+9 \\ -6}$ & $30\substack{+5 \\ -4}$ \\
$e_c$  & $0.15\substack{+0.24 \\ -0.12}$ & $0.20\substack{+0.20 \\ -0.14}$ \\
$i_c$ (°)  & $132\substack{+14 \\ -9}$ & $134\substack{+6 \\ -5}$ \\
$\Omega_c$ (°)  & $301\substack{+34 \\ -164}$ & $330\substack{+16 \\ -161}$ \\
$\omega_c$ (°)  & $192\substack{+91 \\ -147}$ & $200\substack{+48 \\ -140}$  \\
$\theta_c$ (°)  & $270.6\pm0.8$ & $270.3\pm0.6$ \\
\hline
$a_d$ (AU) & $19\substack{+9 \\ -7}$ & $13.2\substack{+1.7 \\ -1.6}$ \\
$e_d$  & $0.16\substack{+0.20 \\ -0.12}$ & $0.14\substack{+0.15 \\ -0.10}$ \\
$i_d$ (°)  & $126\substack{+12 \\ -10}$ & $134\substack{+6 \\ -5}$ \\
$\Omega_d$ (°)  & $166\substack{+79 \\ -127}$ & $330\substack{+16 \\ -161}$ \\
$\omega_d$ (°)  & $196\substack{+95 \\ -129}$ & $211\substack{+66 \\ -113}$ \\
$\theta_d$ (°)  & $293.1\substack{+2.4 \\ -3.0}$ & $290.4\substack{+2.2 \\ -2.0}$ \\
\hline
$M_{\mathrm{sys}}$ ($M_\odot$)  & $0.92\pm0.10$ & $0.90\pm0.09$ \\
$\pi$ (mas)  & $8.898\pm0.020$ & $8.89\pm0.02$ \\
\enddata
\tablecomments{$\theta_b$, $\theta_c$, and $\theta_d$ correspond to the position angle at the reference epoch (MJD~60000). The values listed represent the median and 68\% credible interval for each parameter.}
\end{deluxetable}

\subsection{Updated Orbit Fits and Stability} \label{disc-orbitfits}

\begin{figure*}
    \centering
    \centering{{\includegraphics[width= \textwidth]{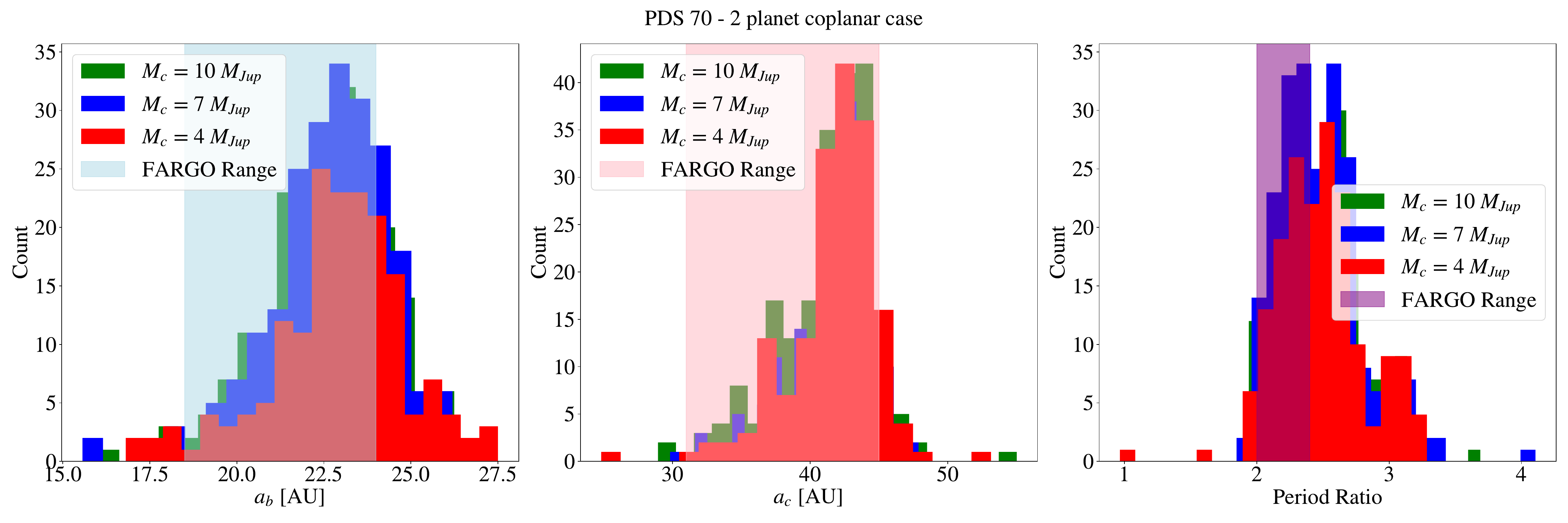} }}%
    \qquad
    \centering{{\includegraphics[width= \textwidth]{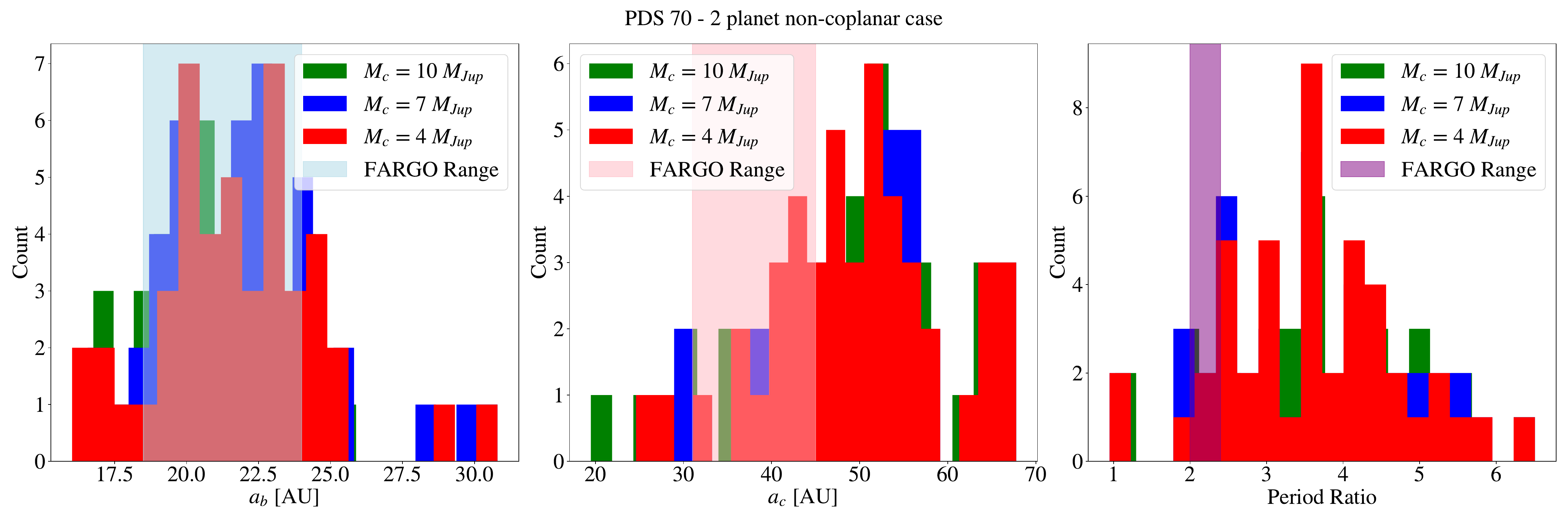} }}%
    \caption{Histograms show the posteriors in the orbit fit that remain stable for 100 Myr timescales. Top: coplanar case, where 4.2 -- 4.5\% of cases are stable. The semi-major axis of b and c (left and middle) lead to a period ratio between $\sim$ 2 -- 3 for both planets (right panel). Bottom: non-coplanar case, where 0.7 -- 0.9\% configurations are stable. The semi-major axis values for b and c lead to stable configurations that favor period ratios between 2 -- 6 (right panel), which is significantly different from our FARGO outputs.}%
    \label{fig:orbitfit-2-stab}
\end{figure*}

We find that the stable configurations from the orbit fits broadly agree with our FARGO outputs, except for the semi-major axis of the outer planet, which is generally found to be at a larger separation from the host star in the orbit fits. Period ratios between 2 -- 2.5 can potentially be in a 2:1 or 5:2 MMR. In the case where the planets are not coplanar, only 0.8 -- 0.9\% or the orbital configurations remain stable for 100 Myr, and they mostly disagree with our FARGO outputs -- the period ratio varies between 2 -- 6, with planet c being at significantly wider separations than in the FARGO case (between 37 -- 67 AU) -- see Figure \ref{fig:orbitfit-2-stab}, bottom row. 
\par
When we include PDS 70 ``d" in the joint orbit fit, using the astrometry listed in \citep{Mesa2019, Christiaens2024}, \textit{none} of the orbit fit configurations remain stable for 100 Myr. This is likely due to the high eccentricity of planet b in the joint orbit fit (see Table \ref{tbl:orbitswithd}). Such high eccentricities can be explained by the small orbital arc coverage due to the long periods of the planets, and will likely change with additional data \citep{DoO2023, DoO2024}. For that reason, we cannot definitively conclude that PDS 70 b is certainly in an eccentric orbit.
\par
Overall, when we do not start from already near-resonant configurations (e.g. from FARGO outputs, which take into account disk migration), the system is very likely to become unstable. This is because the orbit fits are severely undersampled, covering only a few percent of the true orbit, which is evident in Figures \ref{fig:orbitfit-2-stab} and \ref{fig:orbitplots}.  Consequently, the astrometric orbit fits yield large uncertainties in the orbits' parameter space. Since the large masses of the planets likely require near-coplanarity and MMR (and these parameters seem to hold true from planet-disk interaction theory) in order to remain stable, the small region of parameter space where these requirements are met is needed for stability but not favored by blind orbit fits. 
\par

\subsection{Implications for Observations of Mature Systems with Direct Imaging}
Considering the PDS 70 planets' current location in their disk, we evaluated the possible orbital architecture outcomes of the system by varying the planet-disk and planet-planet interactions under the action of photoevaporation, the uncertainties in planet mass ratios and disk viscosity. We found that all of these parameters play a role in how the system will evolve over time, but overall the system is most likely to remain stable. A few factors appear to favor long-term stability for this system, in particular the resonance locking when still in the gas disk and lower planet masses. The planets' migration is mostly minor (i.e. $<$ 2 AU -- the planets are likely to remain around where they are) except in the case where the viscous timescale dominates over the disk lifetime i.e. if the disk is viscous ($\alpha \sim 10^{-2}$) and the disk photoevaporation is small. In that case, the planets migrate outward significantly ($>$ 5 AU) from their current de-projected locations. However, recent works have found that protoplanetary disks are less viscous than previously thought.  For example, \citealt{Pinte2016}'s dust settling modeling of HL Tau found that $\alpha$ of $10^{-4}$ reproduced the disk's observations. Furthermore, the weak accretion of $\sim$ $10^{-10}$ \Msol/yr for PDS 70 \citep{Joyce2023} suggests that the disk's viscosity is indeed smaller than $10^{-2}$ (most likely smaller than $10^{-3}$ as well; see for instance \citealt{Rafikov2017} and \citealt{Portilla-Revelo2023}), at least in its current configuration. \par

Taking into account our results from orbit fits and hydrodynamic simulations, coupled with past works on this system's $\alpha$ and mass measurements, we conclude that the most realistic scenarios are those with $\kappa$ of 0.1 or 1 (from X-ray luminosity and photoevaporation expectations; e.g. \citealt{Joyce2023, Sellek2024}), with $\alpha \simeq 10^{-3}$ (from dust distribution in the PDS 70 disk e.g. \citealt{Bae2019, Portilla-Revelo2023}),
and $M_c < 10 M_{Jup}$ (from planet luminosities, e.g. \citealt{Haffert2019, Close2025}). Therefore, PDS 70 is most likely to be long-term stable given the realistic scenarios from our simulations, and the planets are not expected to migrate significantly in the next 0.2 -- 2 Myr while the disk finishes dissipating. 
\par
Our results suggest that widely separated, super-sized multiplanet gas giant systems can be found at older ages ($>$ Gyr). However, they will only survive under specific circumstances: they must be resonance locked, with low eccentricity. If these specific criteria are not achieved, the system will likely become unstable within a few Myr due to the strong planet-planet interactions. The gas disk is essential to facilitate migration that leads to resonance locking. The photoevaporation, provided that it does not completely halt the migration of the planets, can facilitate the stability by dampening the eccentricities of the planets. Viscosity primarily affects migration pathways and resonance locking; in our simulations, it does not directly destabilize the system.  \par
When PDS 70 d is added to the simulations, the stability of the system is less likely in the long term. \citealt{Nagpal2024} find a similar result in their work: their two-planet system tests cases rarely experienced ejections, which changed significantly once a third planet was present in the system. \par
In this study, we did not model the initial resonance capture process, as our simulations begin with the planets near the observed 2:1 MMR suggested for PDS 70. Instead, we focus on the evolution of the system under disk dispersal after most of the convergent migration has already occurred -- i.e., how we see them in their current configuration. Migration-driven resonance locking is a natural outcome in planet formation models and has been observed in compact multi-giant systems discovered via radial velocities (e.g.,\citealt{Snellgrove2001, AtaieeKley2020, TerquemPapaloizou2007, ChoksiChiang2020}) and proposed for the early Solar System \citep{Griveaud2024}. However, such resonant configurations are rarely seen in older, less massive and compact systems found with transits \citep{Goldreich2014}, suggesting that MMR chains may be disrupted over time for less massive systems. We do not find this result in the PDS 70 case, as the planets mostly remain where they are (near the 2:1 MMR) over long timescales. Notably, the HR 8799 system, which is older than PDS 70 ($\sim$ 30 Myr), shows evidence of long-term resonance stability, with its four planets likely locked in a Laplace resonance chain. This supports the idea that early resonance capture during migration can persist even after disk dispersal in widely separated, highly massive systems. \par
The disk photoevaporation was found to weaken the resonance-locking in all cases tested here, due to its halting of convergent migration. Therefore, disk photoevaporation can play a significant role in the breaking of resonant chains for multiplanet systems. For such massive planets to form and fully migrate into the resonance before the disk dissipates could indicate that they either had a fast formation mechanism that allowed them to form and migrate to the center of the MMR before the disk fully dissipates, or that their natal disks live for at least several Myr, such that a slower formation process and resonance locking can occur. Encouragingly, photoevaporation models that include thermochemistry have found that mass loss rates are lower than previously thought ($<10^{-9}$ \Msol $yr^{-1}$; \citealt{Sellek2024}), which could help explain the highly massive, widely separated and resonant locked gas giants found with direct imaging.

\section{Conclusion} \label{conclusion}
The main findings of this study are: 
\begin{enumerate}
    \item PDS 70 b and c are most likely in resonance, which facilitates long-term stability. The disk's photoevaporation weakens the convergent migration towards resonance locking, consequently weakening the long-term stability of the system.  We find that in every case tested, the system remains stable (regular), without disruption, for $>$ 1 Gyr.
    \item Orbit fit posteriors based on astrometry/RVs only are found to not give accurate long-term assessments of the planets' stability. This is due to the stability requiring a confined parameter space configuration where the planets are in resonance, which is difficult to obtain with highly uncertain posteriors. Additionally, N-body only simulations generally do not account for disk migration, which we found facilitates long-term stability. A possible way to improve current orbit fits (beyond obtaining additional data) is to tightly constrain the parameter space towards near-resonant configurations. Going beyond period ratios, it may be advantageous to use priors on the elements related to the resonant angle to favor a possible libration. This is an important test for future work.
    \item If the third planet PDS 70 ``d" at 13.5 AU is real, it would most likely lock into a Laplace resonance with the outer two planets, consistent with observations, due to the carving of a common gap. We test a case where $M_d$ is 2 $M_{Jup}$, and find that the system can become long-term unstable, with Lyapunov timescales of 6 and 0.004 Myr for $\kappa = \frac{1}{10}$ and 1 respectively. This confirms that the disk's presence stabilizes the system by allowing the planets to migrate towards the center of the MMR. However, we caution that this could change significantly if the masses of the planets are smaller, as was found in the 2-planet case. Since the masses are highly uncertain, it is possible that the system can be long-term stable with a third gas giant planet.
    \end{enumerate}
    
As the direct imaging field moves towards detecting older systems, the dynamical stability of widely separated, highly massive gas giants may strongly affect the occurrence rates of systems with more than one giant. We find here that convergent migration, resonance locking, and low eccentricities play a key role in facilitating this long term stability. In order to confirm these scenarios, it is advantageous to improve detection limits such that mature systems can be probed with direct imaging.

\section{Acknowledgements}
The authors thank Aaron Rosengren, Eve Lee, Sam Hadden and Dan Tamayo for helpful discussions. C.D.O. is supported by the National Science Foundation Graduate Research Fellowship under Grant No. DGE-2038238. Any opinions, findings, and conclusions
or recommendations expressed in this material are those of the author(s) and do not necessarily reflect the views of the National Science Foundation.
K.G. and D.J. acknowledge Poznań Supercomputer Centre (PCSS, grant pl0406-01) and  the Centre of Informatics Tricity Academic Supercomputer and network (CI TASK, Gda\'nsk, Poland) for computing resources (Grant No. PT01016).

\clearpage
\bibliographystyle{aa_bst.bst}
\bibliography{main.bib}
\clearpage
\appendix

\section{Lyapunov Timescale} \label{lyapunov}
\begin{figure*}[hb!]
    \centering
    \centering{{\includegraphics[width= 12cm]{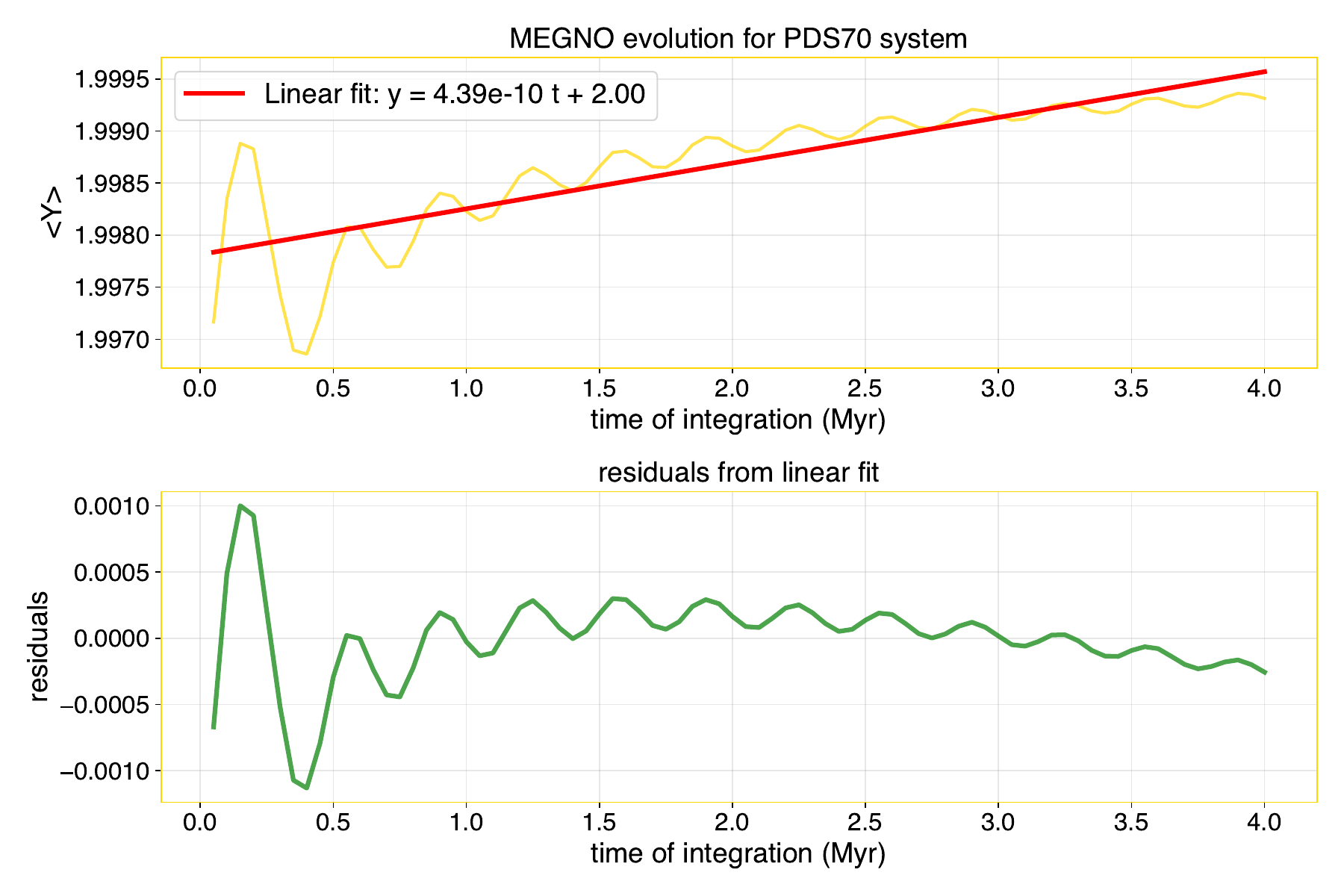} }}%
    \qquad
    \centering {{\includegraphics[width=12cm]{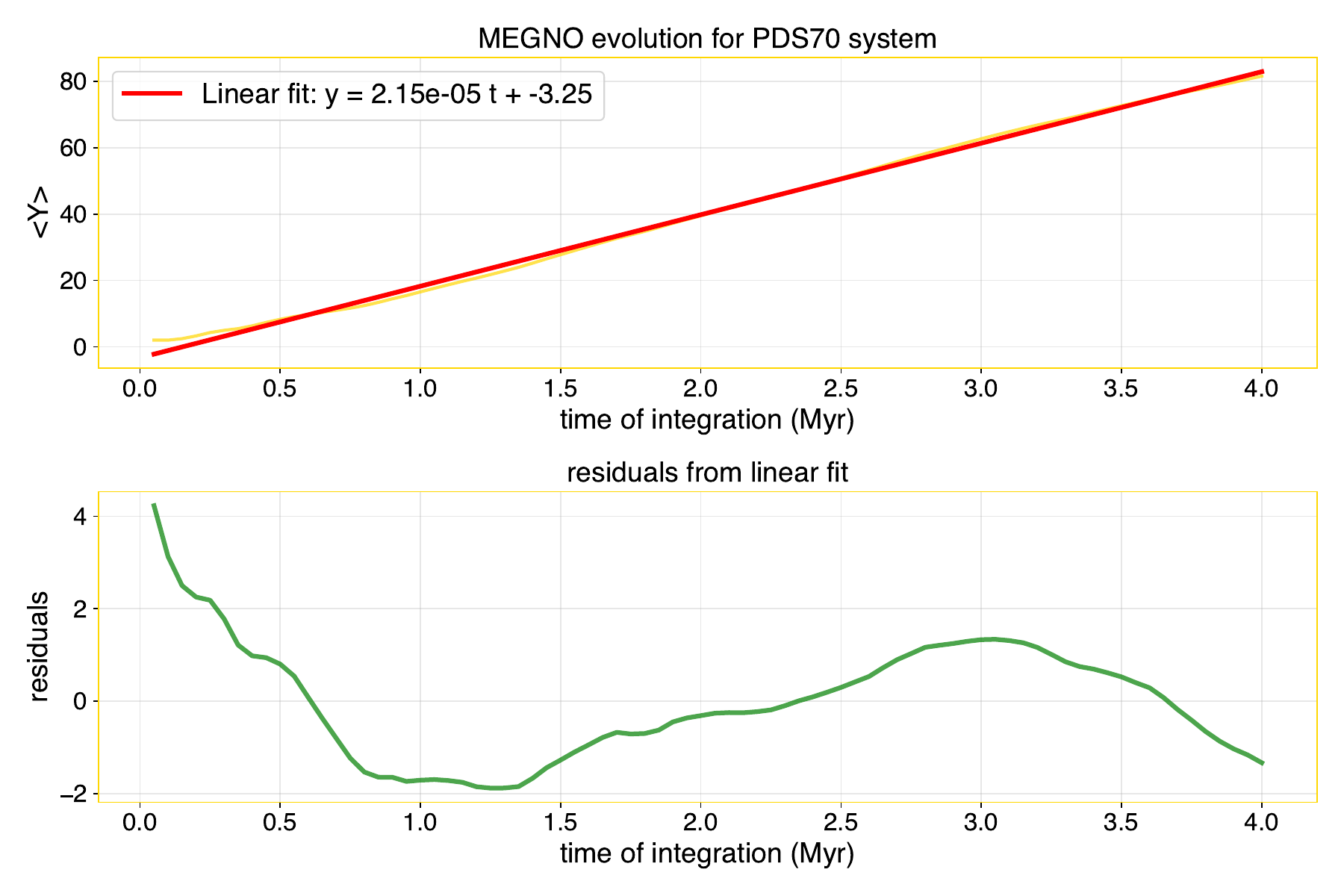} }}
    \caption{Demonstration of the convergence of MEGNO to the value of $\sim$ 2 for a regular (stable) system, which is what we find in our hydrodynamic simulation outputs. For chaotic systems, we can calculate the Lyapunov timescale using the slope of the MEGNO evolution line. We find in this case that increasing the eccentricity of the outer planet by just 0.03 (from 0.06 to 0.09) leads to a chaotic solution, where the Lyapunov timescale is $\sim$ 90,000 years. }%
    \label{fig:lyapunov}
\end{figure*}

Here we present an example on how the MEGNO evolution can be tied to the Lyapunov timescale for chaotic systems \citep[e.g.,][and references therein]{Gozdziewski2001}. We compute the evolution of the MEGNO chaos indicator in REBOUND for 4 Myr equivalent to 20,000 outermost orbits, saving the value every 50,000 yr. We then fit a linear function to the MEGNO aka $\left<Y\right>$ evolution curve, according to the theoretical prediction 
\[
\left<Y\right>(t) \simeq  a t + d,
\]
where, for regular orbits $a=0$ and $d\simeq 2$, while for chaotic (unstable) solutions
the slope determines the maximal Lyapunov exponent (MLCE):
\begin{equation}
    a \sim \frac{1}{2} \lambda.
\end{equation}
Then, the Lyapunov timescale $T_{\rm L}$ is simply the inverse of the MLCE, i.e., $T_{\rm L} = {\lambda}^{-1}$. For a regular system, such as one from the stable region obtained in our hydrodynamic simulations, MEGNO converges to 2 -- yielding a near-zero slope (see Figure \ref{fig:lyapunov}, top panel) and essentially infinite Lyapunov time. Once the system is outside of the stable zone, e.g. once we sufficiently increase the eccentricity of the outer planet, placing it in the MMR overlapping zone, we find an unstable solution. This chaotic solution yields a linear growth in MEGNO, which can be used to obtain the Lyapunov timescale. An example of this is shown in Figure \ref{fig:lyapunov}, where increasing the eccentricity of the outer planet by just 0.03 (from 0.06 to 0.09) leads to a chaotic zone (as seen in Figure \ref{fig:res-structure}) for which the Lyapunov timescale can be calculated. The Lyapunov time $T_{\rm L}$ varies across the parameter space, depending on its local properties and we detected it can be as short as 1000 years. Typically, it can be in the range of a few tens of Kyrs in mildly chaotic regions and much shorter close to the orbit collision zone (Figure \ref{fig:res-structure}). This short $T_{\rm L}$ reflects strong geometric (physical) instability, especially for large outermost masses. We note that determining $T_{\rm L}$ with the linear formulae is much more simple than calculating $\lambda$ directly, from its canonical definition.

\section{Orbit Plots}

\begin{figure*}
    \centering
    \centering{{\includegraphics[width= 8.5cm]{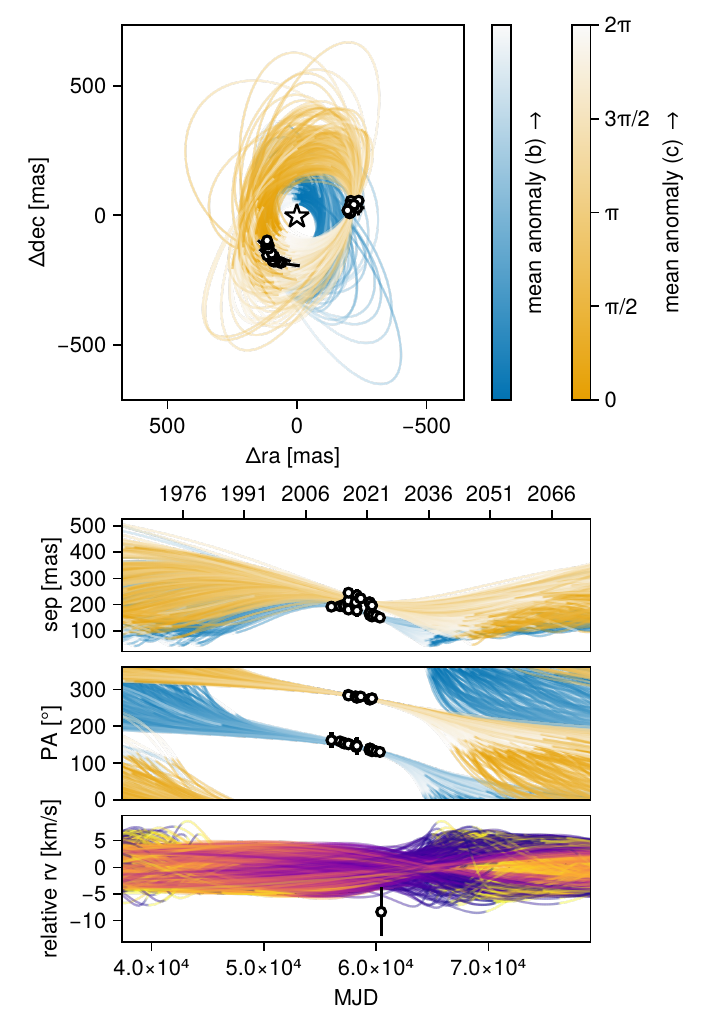} }}%
    \qquad
    \centering {{\includegraphics[width=8.5cm]{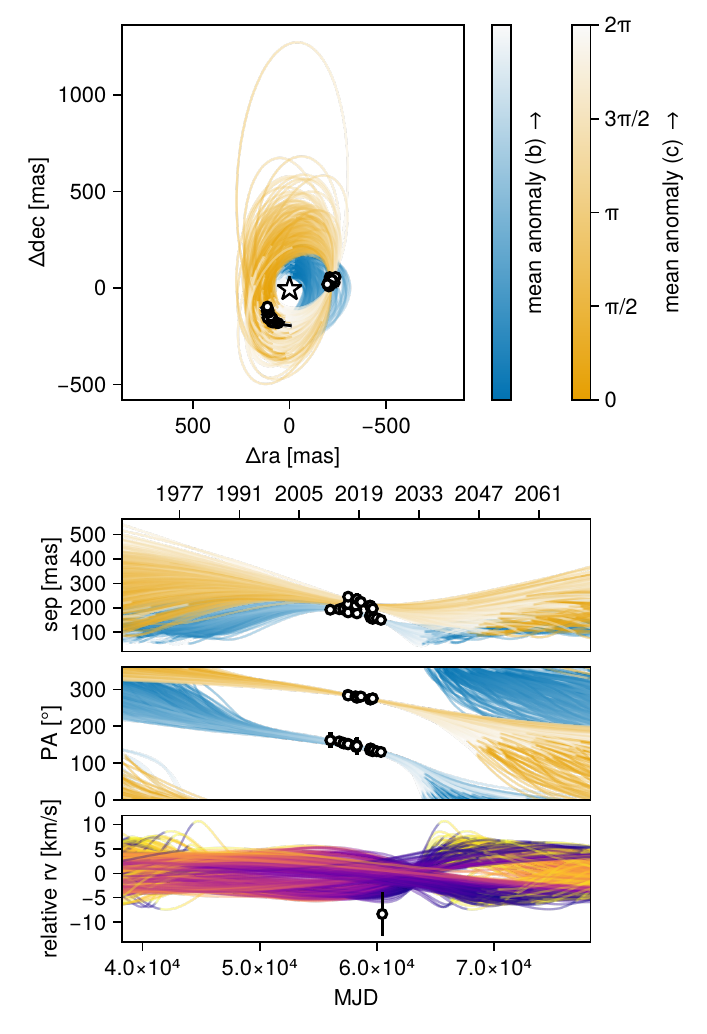} }}
    \caption{Orbit plots for the unconstrained (left) and coplanar (right) for PDS 70 b and c. The top image traces 100 possible orbits on the plane of the sky, while the bottom images' panels represent the separation (sep; in mas), position angle (PA; in degrees) and relative radial velocities (RVs; in km/s) as a function of epoch (in MJD) for both planets. Planet b's fits are shown in orange while planet c's fits are shown in blue.}%
    \label{fig:orbitplots}
\end{figure*}

\begin{figure*}
    \centering
    \centering{{\includegraphics[width= 8.5cm]{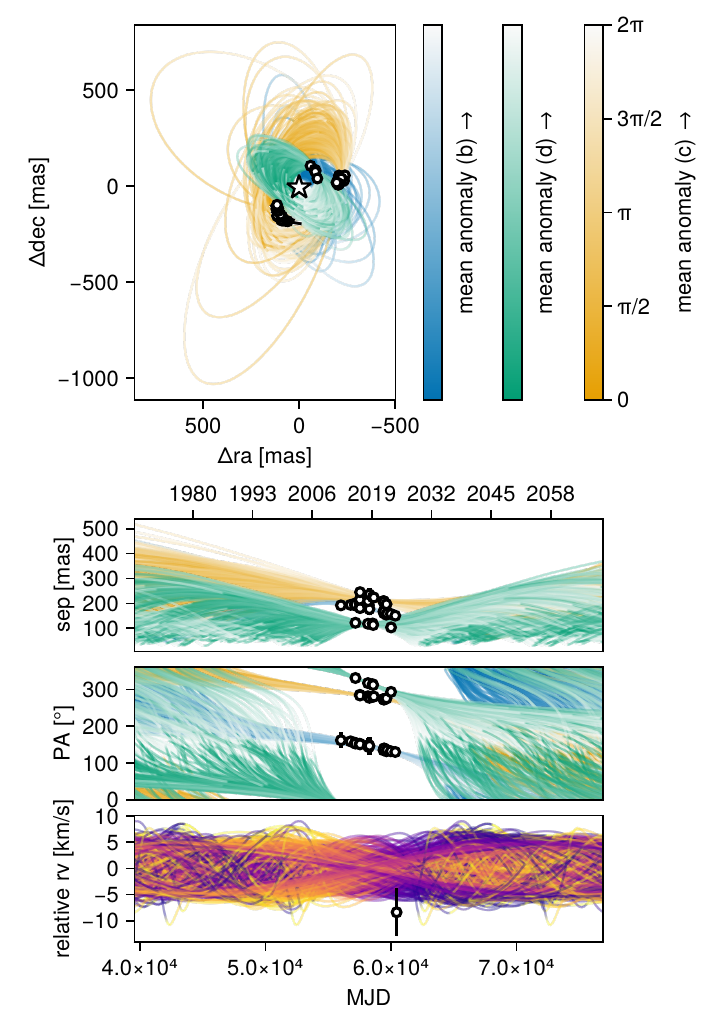} }}%
    \qquad
    \centering {{\includegraphics[width=8.5cm]{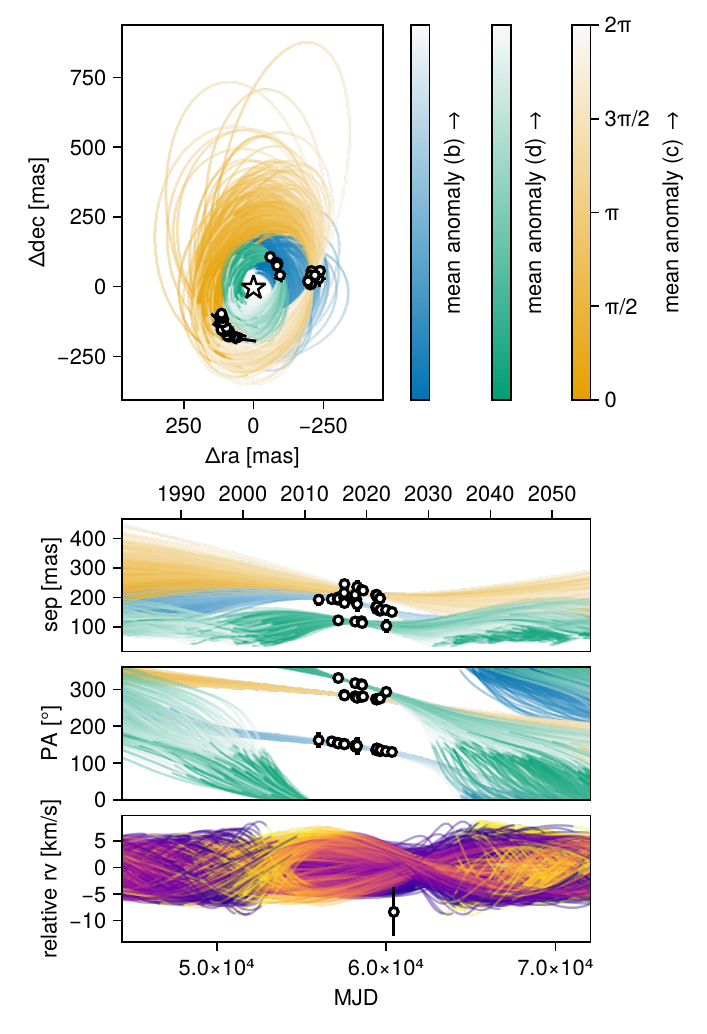} }}
    \caption{Orbit plots for the unconstrained (left) and coplanar (right) for PDS 70 b, c and d. The top image traces 100 possible orbits on the plane of the sky, while the bottom images' panels represent the separation (sep; in mas), position angle (PA; in degrees) and relative radial velocities (RVs; in km/s) as a function of epoch (in MJD) for both planets. Planet b's fits are shown in orange while planet c's fits are shown in blue and planet d's fits are shown in green.}%
    \label{fig:orbitplots-withd}
\end{figure*}

We present the orbit plots for the two and three-planet case for our orbit fits from section \ref{disc-orbitfits}, including the unconstrained and coplanar cases in each panel. The two-planet case can be found in Figure \ref{fig:orbitplots} and the three-planet case can be found in Figure \ref{fig:orbitplots-withd}.

\section{2D Snapshots of PDS 70}
Here we present 2D Snapshots for different $\kappa$ values at t $\sim$ 0.6 Myr of integration. It is evident from the plots that the disk efficiently dissipates around the planets (Hill radii marked in yellow) for higher $\kappa$.
 \begin{figure*}
    \centering
    \centering{{\includegraphics[width=\textwidth]{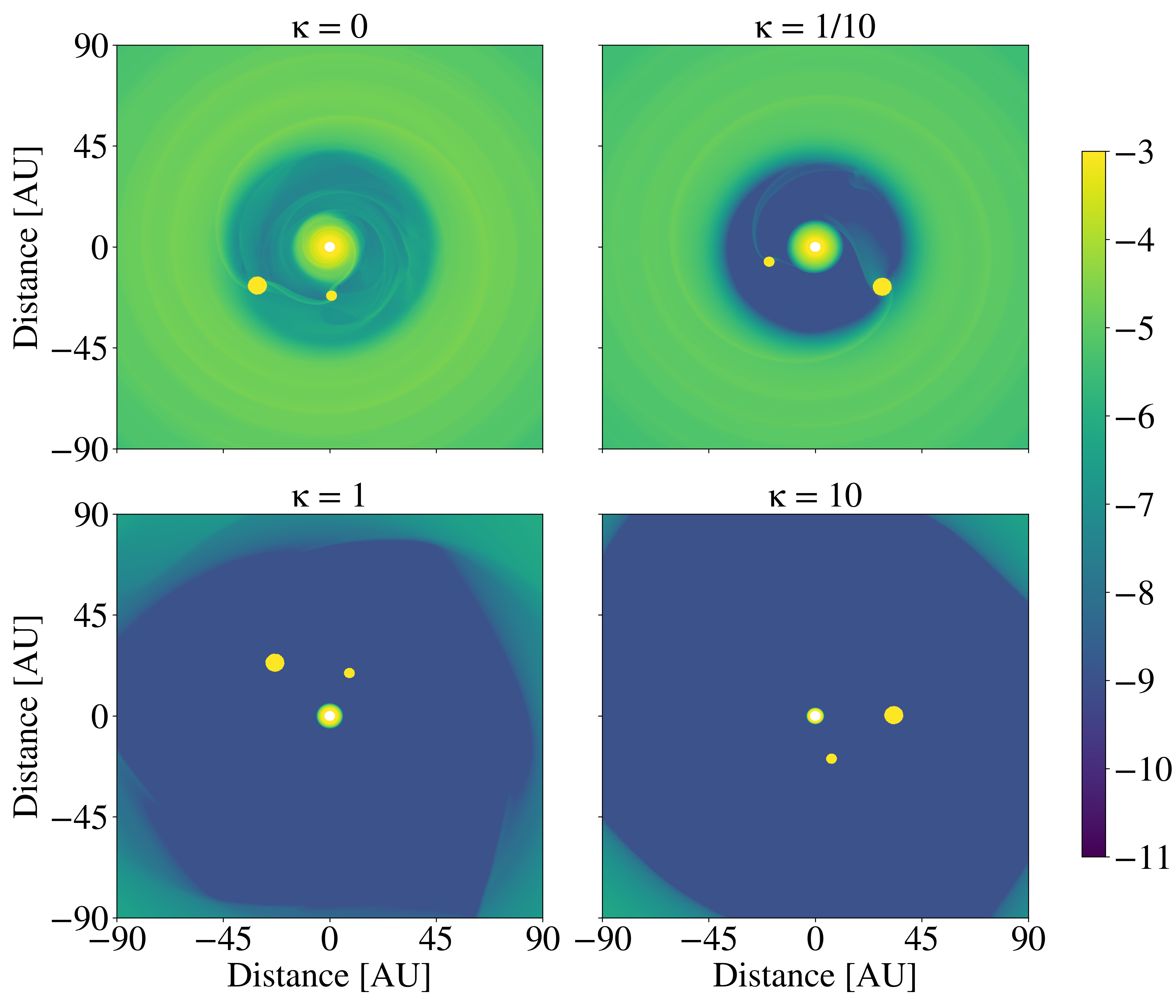} }}
    \caption{Comparison of the 2D density for the disk at t$\sim$ 0.6 Myr for different $\kappa$ values. The colorbar is in units of $\log_{10}(\Sigma_{\mathrm{gas}})\,[M_\odot\,\mathrm{AU}^{-2}]$. The planets' Hill ``circles" are marked in yellow.}%
    \label{fig:2dsnapshot}
\end{figure*}

\end{document}